\documentclass[twoside]{IEEEtran}
\usepackage[utf8]{inputenc}

\usepackage{amsmath}
\usepackage{amsfonts}
\usepackage{amssymb}
\usepackage{cite}

\usepackage{amsthm}
\usepackage{multirow}

\usepackage{color}
\usepackage{mathtools}
\usepackage{xparse}
\DeclarePairedDelimiterX{\set}[1]{\{}{\}}{\setargs{#1}}
\NewDocumentCommand{\setargs}{>{\SplitArgument{1}{;}}m}
{\setargsaux#1}
\NewDocumentCommand{\setargsaux}{mm}
{\IfNoValueTF{#2}{#1} {#1\,\delimsize|\,\mathopen{}#2}}%{#1\:;\:#2}

\DeclarePairedDelimiter\floor{\lfloor}{\rfloor}
\DeclarePairedDelimiter\parenv{\lparen}{\rparen}

\newtheorem{theorem}{Theorem$\!$}

\newtheorem{lemma}{Lemma$\!$}

\newtheorem{corollary}{Corollary$\!$}
\newtheorem{definition}{Definition$\!$}

\newtheorem{remark}{Remark$\!$}
\newtheorem{example}[theorem]{Example$\!$}

\newcommand{\cN}{\mathcal{N}}

\newcommand{\cS}{\mathcal{S}}
\newcommand{\cT}{\mathcal{T}}

\renewcommand{\leq}{\leqslant}
\renewcommand{\geq}{\geqslant}

\newcommand{\F}{\mathbb{F}}
\newcommand{\Fq}{\F_{\!q}}
\newcommand{\eps}{\varepsilon}

\newcommand{\sbinom}[2]{\genfrac{[}{]}{0pt}{}{#1}{#2}}

\DeclareMathOperator{\rank}{rank}
\DeclareMathOperator{\gap}{gap_2}
\usepackage{pgf}
\usepackage{tikz}
\usetikzlibrary{matrix,chains,positioning,arrows,calc,decorations, plotmarks, patterns, fit,backgrounds}
\usetikzlibrary{arrows.meta,matrix,positioning}
\usepackage{pgfplots}
\usetikzlibrary{external,fillbetween,svg.path}
\pgfplotsset{compat=1.3}
\tikzstyle{help lines}=[black!20,dashed]

\pgfplotscreateplotcyclelist{mylist}{red,blue,black,yellow,brown}

\pgfdeclarepatternformonly{my north east lines}{\pgfqpoint{-1pt}{-1pt}}{\pgfqpoint{8pt}{8pt}}{\pgfqpoint{6pt}{6pt}}%
{
	\pgfsetlinewidth{0.4pt}
	\pgfpathmoveto{\pgfqpoint{0pt}{0pt}}
	\pgfpathlineto{\pgfqpoint{6.1pt}{6.1pt}}
	\pgfusepath{stroke}
}
\definecolor{light_gray}{rgb}{0.6,0.6,0.6}
\definecolor{awgray}{rgb}{0.7,0.7,0.7}
\definecolor{awgray_dark}{rgb} {0.4,0.4,0.4}
\tikzset{
	>=stealth',
	mycircle/.style={circle, draw=gray, very thick, text width=.1em, minimum height=1.5em, text centered},
	mycircle_small/.style={circle,draw=awgray_dark,fill = awgray_dark, inner sep=0,minimum size=.6em},
	mycircle_small_black/.style={circle,draw=black,fill = black, inner sep=0,minimum size=.6em},
	mybox/.style={rectangle,rounded corners,draw=black, thick,text width=1em,minimum height=4em,minimum width=4em,text centered},
	mybox_small/.style={rectangle,rounded corners,draw=black, thick,text width=1em,minimum height=2em,minimum width=2em,text centered},
	mybox_vec/.style={rectangle,rounded corners,draw=black, thick,text width=1em,minimum height=0.7em, minimum width=4em,text centered},
	mybox_vec_short/.style={rectangle,rounded corners,draw=black, thick,text width=1em,minimum height=0.7em, minimum width=2em,text centered},
	pil/.style={->, thick, shorten <=2pt, shorten >=2pt,},
      }

\def\ve#1{{\mathchoice{\mbox{\boldmath$\displaystyle #1$}}%
              {\mbox{\boldmath$\textstyle #1$}}%
              {\mbox{\boldmath$\scriptstyle #1$}}%
              {\mbox{\boldmath$\scriptscriptstyle #1$}}}}

\newcommand{\A}{\ve{A}}
\newcommand{\B}{\ve{B}}
\newcommand{\x}{\ve{x}}
\newcommand{\y}{\ve{y}}

\newcommand{\quadbinom}[2]{\sbinom{#1}{#2}}

\usepackage{xcolor}
\definecolor{brightmaroon}{rgb}{0.76, 0.13, 0.28}
\definecolor{ao}{rgb}{0.0, 0.5, 0.0}
\definecolor{azure}{rgb}{0.0, 0.5, 1.0}
\definecolor{TUMBlue}{RGB}{0,101,189} % Pantone 300
\definecolor{TUMBlueDark}{RGB}{0,82,147} % Pantone 301
\definecolor{TUMBlueLight}{RGB}{152,198,234} % Pantone 283
\definecolor{TUMBlueMiddle}{RGB}{100,160,200} % Pantone 542

\definecolor{TUMElfenbein}{RGB}{218,215,203} % Pantone 7527 -Elfenbein
\definecolor{TUMGreen}{RGB}{162,173,0} % Pantone 383 - Grün
\definecolor{TUMOrange}{RGB}{227,114,34} % Pantone 158 - Orange
\definecolor{TUMGray}{gray}{0.6} % Grau 60%

\definecolor{TUMGreenLight}{RGB}{0,124,48}
\definecolor{TUMRed}{RGB}{196,7,27}
\begin{document}

%----------------- The Title Declarations ------------------------------

\IEEEoverridecommandlockouts
\title{On the Gap between Scalar and Vector Solutions of Generalized Combination Networks}

\author{
    Hedongliang Liu,~\IEEEmembership{Student Member,~IEEE},
    Hengjia Wei,
    Sven Puchinger,~\IEEEmembership{Member,~IEEE}\\
    Antonia Wachter-Zeh,~\IEEEmembership{Senior Member,~IEEE},
    Moshe Schwartz,~\IEEEmembership{Senior Member,~IEEE}%
    \thanks{This research was supported in part by a German Israeli Project Cooperation (DIP) grant under grant no.~PE2398/1-1 and KR3517/9-1, and by the European Union's Horizon 2020 research and innovation programme under the Marie Sk\l{}odowska-Curie grant agreement no.~713683. The material in this paper was partly presented at the 2020 International Symposium on Information Theory (ISIT).}%
    \thanks{Hedongliang Liu, Sven Puchinger, and Antonia Wachter-Zeh are with the Department of Electrical and Computer Engineering, Technical University of Munich, Munich 80333, Germany (e-mail: lia.liu@tum.de; sven.puchinger@tum.de; antonia.wachter-zeh@tum.de). This work was partly done while Sven Puchinger was with the Department of Applied Mathematics and Computer Science, Technical University of Denmark, Lyngby, Denmark.}%
    \thanks{Hengjia Wei and Moshe Schwartz is with the School of Electrical and Computer Engineering, Ben-Gurion University of the Negev, Beer Sheva 8410501, Israel (e-mail: hjwei05@gmail.com; schwartz@ee.bgu.ac.il)}%
    % \thanks{Sven Puchinger is with the Department of Electrical and Computer Engineering, Technical University of Munich, Munich 80333, Germany (e-mail: sven.puchinger@tum.de). This work was partly done while Sven Puchinger was with the Department of Applied Mathematics and Computer Science, Technical University of Denmark, Lyngby, Denmark.}%
    % \thanks{Moshe Schwartz is with the School of Electrical and Computer Engineering, Ben-Gurion University of the Negev, Beer Sheva 8410501, Israel (e-mail: )}%
}

% The paper headers % commented for arXiv version
% \markboth{IEEE TRANSACTIONS ON INFORMATION THEORY,~Vol.~X, No.~X, MONTH~YEAR}%
% {Liu \MakeLowercase{\textit{et al.}}: {On the Gap between Scalar and Vector Solutions of Generalized Combination Networks}}

%% % make the title area
\maketitle
\begin{abstract}
We study scalar-linear and vector-linear solutions of the generalized combination network. We derive new upper and lower bounds on the maximum number of nodes in the middle layer, depending on the network parameters and the alphabet size. These bounds improve and extend the parameter range of known bounds. Using these new bounds we present a lower bound and an upper bound on the gap in the alphabet size between optimal scalar-linear and optimal vector-linear network coding solutions. For a fixed network structure, while varying the number of middle-layer nodes $r$, the asymptotic behavior of the upper and lower bounds shows that the gap is in $\Theta(\log(r))$.
\end{abstract}


%%%%%%%%%%%%%%%%%%%%%%%%%%%%%%%%%%%%%%%%%%%%%%%%%%%%%%%%%%%%%%%
%%%%%%%%%%%%%%%%%%%%%%%%%%%%%%%%%%%%%%%%%%%%%%%%%%%%%%%%%%%%%%%
%%%%%%%%%%%%%%%%%%%%%%%%%%%%%%%%%%%%%%%%%%%%%%%%%%%%%%%%%%%%%%%

\section{Introduction}
\label{sec:intro}
\IEEEPARstart{I}{n} multicast networks that apply routing, a source node multicasts information to other nodes in the network in a multihop fashion, where every node can pass on their received data. \emph{Network coding} has been attracting increasing attention since the seminal papers~\cite{ACLY00,LYCfeb2003} {which showed that the throughput can be increased significantly by not just forwarding packets but also performing linear combinations of them}. Several follow-up works~\cite{NYoct2004,HJsep2009,MLLfeb2012,EGLapr2014} also showed that network coding outperforms routing in terms of delay, throughput and reliability for specific networks.

In network coding, each node is allowed to encode its received data before passing it on. We formulate the \emph{network coding problem} as follows: for each node in the network, find a function of its incoming messages to transmit on its outgoing links, such that each receiver can recover all (or a predefined subset of all) the messages. We say a network is \emph{solvable} if such a function exists. %  \AW{[TODO: Define what solvable means]}
The encoding at relay nodes incurs delay and memory cost in the network. One approach in minimizing these costs, is reducing the alphabet size of the coding operations, thus resulting in less complexity in practical implementations of network coding~\cite{LSB2006,LS2009,GSRMsep2019}.

\subsection{Previous Work}
A considerable number of studies have been conducted on different types of network coding:
such as linear network coding~\cite{LYCfeb2003,KM03} and non-linear network coding~\cite{LLjan2004}, deterministic network coding~\cite{XiaMedAul07} and random linear network coding~\cite{KK08,SKK08}. In this paper, we only focus on linear network coding and discuss the performance of scalar linear network coding and vector linear network coding.

  In linear network coding, each linear function for a receiver consists of coding coefficients for incoming messages. If the messages are scalars in $\Fq$ and the coding coefficients are vectors over $\Fq$, the solution is called a scalar linear solution. If the messages are vectors in $\Fq^t$, and the coding coefficients are matrices over $\Fq$, it is called a vector linear solution. Vector network coding was mentioned in~\cite{CDFZ06} as fractional network coding and extended to vector network coding in~\cite{EFfeb2011}.

  Although a scalar solution over $\F_{q^t}$ can be translated to a vector solution composed of $t\times t$ matrices over $\Fq$, directly designing codes for vector network coding still has advantages: there exist $q^{t^2}$ many $t\times t$ matrices over $\Fq$, while a scalar solution only employs $q^t$ of them. Therefore, vector network coding offers a larger space of choices for optimizing the performance of a network. However, not every solvable network has a vector solution~\cite{DFZaug2005}. The hardness of finding a capacity-achieving vector solution for a general instance of the network coding problem was proved in~\cite{LSfeb2011}. In~\cite{DRsep2016} it was proved that {for a class of non-multicast networks, a vector linear solution of dimension $t$ exists but no vector solution over any finite field exists if the message dimension is less than $t$.}
  {The existence of explicit networks where scalar solutions still outperform binary vector solutions was shown in~\cite{SYL+dec2016}.}
{Nevertheless, a {multicast} network was constructed in~\cite{SYL+dec2016} whose minimal alphabet for a scalar linear solution is strictly larger than the minimal alphabet for a vector linear solution.}
The gap in the minimum alphabet size between a scalar solution and a vector solution was shown to be positive in generalized combination networks~\cite{EW18} and minimal multicast networks~\cite{CCESW20}. Several algorithms for deterministic networks via vector coding were presented in~\cite{EFjan2010,EFjun2010,EFfeb2011,EW18}.

  Solving network coding problems also motivates research in other topics such as new metrics for network codes~\cite{EZjul2019}, subspace codes design~\cite{GYmay2010,ESfeb2013,EKOOfeb2020}, networks over the erasure channel~\cite{GBFTjul2011} and distributed storage~\cite{DGW+sep2010,DRWS2011}.
More long-standing open problems can be found in~\cite{FS2016}.

\subsection{Our Goals and Contributions}
In this paper, we only consider linear solutions of {multicast} networks.
Denote by $\Fq$ a finite field of size $q$. Bold lowercase letters denote vectors and bold capital letters denote matrices.

The scalar and vector solutions stand for scalar linear and vector linear solutions throughout the rest of the paper.
We call a scalar solution over $\Fq$ for a network, a $(q,1)$-{linear solution}, and we call a {vector solution} of length $t$ over $\Fq$, a $(q,t)$-linear solution.

\begin{figure}[t!]
  \centering
  \def\x{0.55}%0.05\columnwidth} % define a scaling factor

\begin{tikzpicture}
  %%%  tikz predefined styles
  [font=\normalsize,>=stealth',
  mycircle/.style={circle, draw=TUMGray, very thick, text width=.1em, minimum height=1.5em, text centered},
  mycircle_small/.style={circle,draw=TUMGray!90,very thick, inner sep=0,minimum size=1em,text centered},
  mylink/.style={color=TUMBlueLight, thick},
  mylink_dir/.style={color=TUMGreen, thick, dashed}]

  %%% Source node
  \coordinate (Source) at (0*\x,4*\x);
  {\node[mycircle,label=above right:{$\ve{x}=({x}_1,{x}_2,\dots,{x}_h)$}] (Source) {};}
%   {\node[mycircle,label=right:{$\ve{x}=\textcolor{TUMBlueDark}{(\ve{x}_1,\ve{x}_2,\dots,\ve{x}_h)\in\Fq^{ht}}$}] (Source) {};}

  %%% Middle nodes
  \node[mycircle_small,below left = \x*60pt and \x*100pt of Source] (M0) {};
  \node[mycircle_small,right = \x*20pt of M0] (M1) {};
  \node[mycircle_small,right = \x*20pt of M1] (M2) {};
  \node[mycircle_small,right = \x*20pt of M2] (M3) {};
  \node[draw=none,right = \x*20pt of M3] (M4) {$\dots$};
  \node[mycircle_small,right = \x*20pt of M4] (M5) {};
  \node[mycircle_small,right = \x*20pt of M5] (M6) {};

  %%% receivers
  \node[mycircle,below left = \x*60pt and \x*20pt of M0,label=below:{$\ve{y}_1$}] (R0) {};
  \node[mycircle,right = \x*50pt of R0,label=below:{$\ve{y}_2$}] (R1) {};
  \node[mycircle,right = \x*50pt of R1,label=below:{$\ve{y}_3$}] (R2) {};
  \node[draw=none,right = \x*50pt of R2] (R3) {$\dots$};
  \node[mycircle,right = \x*50pt of R3,label=below:{$\ve{y}_{N}$}] (R4) {};

  %%% connection between source and middle layer
  \path[] (Source) edge[mylink,bend right=15] (M0.north)
  edge[mylink,bend right=15] (M1.north)
  edge[mylink,bend right=15] (M2.north)
  edge[mylink,bend right=15] (M3.north)
  edge[mylink,bend left=15] (M5.north)
  edge[mylink,bend left=15] (M6.north);

  %%% connection between middle layer and receivers
  \path[] (R0) edge[mylink,bend left=15] (M0.south)
  edge[mylink,bend left=15] (M1.south);
  \path[] (R1) edge[mylink,bend left=15] (M0.south)
  edge[mylink,bend left=15] (M2.south);
  \path[] (R2) edge[mylink,bend left=15] (M1.south)
  edge[mylink,bend left=15] (M3.south);
  \path[] (R4) edge[mylink,bend right=15] (M5.south)
  edge[mylink,bend right=15] (M6.south);

  %%% variable notations
  \draw [mylink,solid] ($(M0)+(\x*18pt,\x*15pt)$) arc (30:80:\x*15pt); % the arc of \ell
  \draw [mylink,-] ($(R0)+(\x*20pt,\x*10pt)$) arc (0:100:\x*15pt); % the arc of \alpha\ell
  \node[draw=none,above right = \x*-8pt and \x*2pt of M0] (ell) {\textcolor{TUMBlue}{$\ell$}};
  \node[draw=none,above right=0pt and -5pt of M6] (rs) {$r$ middle nodes};
  \node[draw=none,right=0pt of R4] (N) {$N=\binom{r}{\alpha}$};
  \node[draw=none,below right=\x*-5pt and \x*-75pt of N] (recNodes) {receivers};

  %%% generalized links
  %%% parallel links
  \path[] (Source) edge[mylink,bend right=5] (M0.north)
  edge[mylink,bend right=5] (M1.north)
  edge[mylink,bend right=5] (M2.north)
  edge[mylink,bend right=5] (M3.north)
  edge[mylink,bend left=5] (M5.north)
  edge[mylink,bend left=5] (M6.north);

  \path[] (R0) edge[mylink,bend left=5] (M0.south)
  edge[mylink,bend left=5] (M1.south);
  \path[] (R1) edge[mylink,bend left=5] (M0.south)
  edge[mylink,bend left=5] (M2.south);
  \path[] (R2) edge[mylink,bend left=5] (M1.south)
  edge[mylink,bend left=5] (M3.south);
  \path[] (R4) edge[mylink,bend right=5] (M5.south)
  edge[mylink,bend right=5] (M6.south);

  %%% direct links
  \path[] (Source.west) edge[mylink_dir, bend right=45] (R0.west)
  edge[mylink_dir, bend right=50] (R0.west)
  (Source) edge[mylink_dir,bend left=5] (R1.east)
  edge[mylink_dir,bend left=10] (R1.east)
  (Source) edge[mylink_dir,bend left=15] (R2.east)
  edge[mylink_dir,bend left=20] (R2.east)
  (Source) edge[mylink_dir,bend right=0] (R4.west)
  edge[mylink_dir,bend right=5] (R4.west);

  %%% notations
  \draw [mylink_dir,solid] ($(R0)+(-\x*15pt,\x*35pt)$) arc (100:60:\x*15pt); % the arc of \epsilon
  \node[draw=none,above left = \x*10pt and 3pt of R0] (alphal) {\textcolor{TUMGreen}{$\varepsilon$}};
  \node[draw=none,right = 0pt of R0] (alphal) {\textcolor{TUMBlue}{$\alpha\ell$}};
  % \node[draw=none,right=0pt of M6] (rv) {$r_v$ middle nodes};

\end{tikzpicture}

%%% Local Variables:
%%% mode: latex
%%% TeX-master: "Journal_final"
%%% End:
  \caption{Illustration of $(\eps,\ell)-\mathcal{N}_{h,r,\alpha\ell+\eps}$ networks}
  \label{fig:Network}
\end{figure}
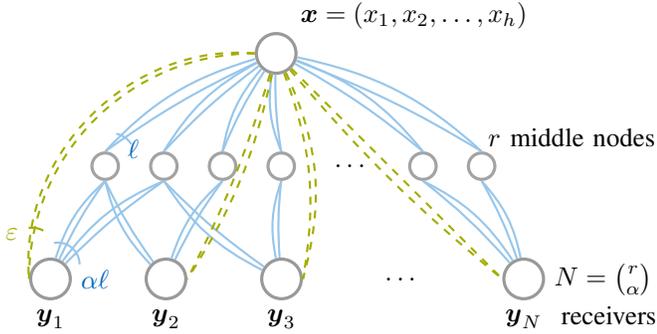

The main object we study in this paper is the class of \emph{generalized combination networks}. An $(\eps,\ell)-\cN_{h,r,\alpha\ell+\eps}$ generalized combination network is illustrated in Figure~\ref{fig:Network} (see also~\cite{EW18}).
The network has three layers. The first layer consists of a source with $h$ source messages. The source transmits $h$ messages to $r$ middle nodes via $\ell$ parallel links (solid lines) between itself and each middle node. Any $\alpha$ middle nodes in the second layer are connected to a unique receiver (again, by $\ell$ parallel links each). Each receiver is also connected to the source via $\eps$ direct links (dashed lines).
It was shown in~\cite[Thm.~8]{EW18} that the $(\eps,\ell)-\cN_{h,r,\alpha\ell+\eps}$ network has a trivial solution if $h\leq\ell+\eps$ and it has no solution if $h>\alpha\ell+\eps$. In this paper we focus on non-trivially solvable networks, so it is assumed $\ell+\eps< h \leq \alpha\ell+\eps$ throughout the paper.

The goal of this paper is to investigate {the maximum number of middle-layer nodes, denoted by $r_{\max}$, such that the network with fixed $h,\alpha,\ell,\eps$ has a $(q,t)$-linear solution. This implies bounds on} the gap between the minimum required alphabet size for scalar and vector solutions of generalized combination networks.
In order to derive the gap size, a metric to measure the improvement has to be specified.
We follow the notations from~\cite{CCESW20} to distinguish between optimal scalar and vector solutions. Given a generalized combination network $\cN$, let $$q_s(\cN):=\min\set*{q ; \cN\textrm{ has a } (q,1)\textrm{-linear solution}}.$$
The $(q_s(\cN),1)$-linear solution is said to be \emph{scalar-optimal}.
Similarly, let $$q_v(\cN):=\min\set*{q^t ; \cN\textrm{ has a } (q,t)\textrm{-linear solution}}.$$
Note that $q_v(\cN)$ is defined by the size of the vector space, rather than the field size.
For $q^t = q_v(\cN)$, a $(q,t)$-linear solution is called \emph{vector-optimal}.
By definition,
\[
q_s(\cN)\geq q_v(\cN).
\]
We define the \emph{gap} as
\[ \gap (\cN):= \log_2 (q_s(\cN)) -\log_2(q_v(\cN)),\]
which intuitively measures the advantage of vector network coding by the amount of extra bits per transmitted symbol we have to pay for an optimal scalar-linear solution compared to an optimal vector-linear solution. {We note that although the definition of the gap differs from the definition of gap in~\cite{CCESW20}, it has been implicitly used in~\cite{EW18}, and mentioned as \emph{information gap} in~\cite{CCESW20}.}

Our main contributions are the following:
\begin{itemize}

    \item two upper bounds on $r_{\max}$, the maximal number of nodes in the middle layer of a generalized combination network {such that the network has a $(q,t)$-linear solution} (Corollary~\ref{cor:imupperbound-N}, valid only for $h\geq 2\ell+\eps$, and Corollary~\ref{cor:imupperbound-2-Network} for $\alpha=2$),
    \item two lower bounds on $r_{\max}$ (Theorem~\ref{thm:LLL_bound} and Corollary~\ref{cor:EK19_lb} for $h\leq 2\ell+\eps$),
    \item an upper bound on the gap in the minimum alphabet size for any fixed generalized combination network structure (Theorem~\ref{thm:gap_ub}),
    \item a lower bound on the gap (Theorem~\ref{thm:gap_lb}) .

\end{itemize}

Our new upper bound on $r_{\max}$ is better than a previous bound from~\cite{EZjul2019} (recalled in Corollary~\ref{cor:EZ19_vector}) for $h\geq 2\ell+\eps$, and the lower bounds outperform previous ones for the whole parameter range of non-trivially solvable generalized combination networks, and they agree with our upper bound up to a small constant factor, for $h\leq 2\ell$ or $h\geq 2\ell,\alpha=2$.

To the best of our knowledge, our upper and lower bounds on the gap are the first such bounds considering fixed network parameters. These bounds are valid for all generalized combination networks with $\eps\neq 0$. The asymptotic behavior of the upper and lower bound shows that $\gap(\cN) = \Theta\parenv*{\log(r)}$.

\subsection{Paper Organization}
The rest of this paper is organized as follows.
In Section~\ref{sec:ub_rs}, we present two new upper bounds on the maximum number of middle-layer nodes, and in Section~\ref{sec:lb_rv} we give two new lower bounds on it. In Section~\ref{sec:bound_gap} we show the gap between the field sizes of scalar-linear and vector-linear solutions. In Section~\ref{sec:discussion}, we compare our upper and lower bounds on the maximum number of nodes in the middle layer with the other known bounds.

\section{Upper Bounds on The Maximum Number of Middle Layer Nodes}
\label{sec:ub_rs}

In this section we fix the network parameters $\alpha,\ell,\eps, h$ and we bound from above the {maximum} number of nodes in the middle layer {such that the network has a $(q,t)$-linear solution}. The main result is given in Corollary~\ref{cor:imupperbound-N} and Corollary~\ref{cor:imupperbound-2-Network}.

We denote by $\mathcal{G}(n,k)$ the Grassmannian of dimension $k$, which is a set of all $k$-dimensional subspaces of $\mathbb{F}_q^n$. The cardinality of $\mathcal{G}(n,k)$ is the well-known $q$-binomial coefficient (a.k.a.~the Gaussian coefficient):
\begin{equation*}
  |\mathcal{G}(n,k)|=\quadbinom{n}{k}_q := \prod\limits_{i=0}^{k-1} \frac{q^n-q^i}{q^k-q^i}=\prod\limits_{i=0}^{k-1} \frac{q^{n-i}-1}{q^{k-i}-1}.
\end{equation*}
A good approximation of the $q$-binomial coefficient can be found in~\cite[Lemma 4]{KK08}:
\begin{equation}
    \label{eq:gauss}
q^{k(n-k)}\leq \quadbinom{n}{k}_q < \gamma \cdot q^{k(n-k)},
\end{equation}
where $\gamma\approx 3.48$.

\begin{lemma}\label{lm-full-t2}
Let $\alpha\geq 2$, $h,\ell,t\geq 1$, $\eps\geq 0$, $h-\eps \geq 2 \ell$, and let $\cT$ be a collection of subspaces  of $\F_q^{(h-\eps) t}$ such that
\begin{enumerate}
\item[(i)] each subspace has dimension at most $\ell t$; and
\item[(ii)] any subset of $\alpha$ subspaces spans $\F_q^{(h-\eps)t}$.
\end{enumerate}
Then  we have $\alpha \ell  \geq h-\eps$ and
$$|\cT| \leq \parenv*{\floor*{ \frac{h-\eps}{\ell} }-2} +\parenv*{\alpha- \floor*{ \frac{h-\eps}{\ell} }+1} \sbinom{\ell t+1}{1}_q.$$
\end{lemma}

\begin{IEEEproof}
Take arbitrarily $\lfloor \frac{h-\eps}{\ell} \rfloor-2$ subspaces  from $\cT$ and take arbitrarily a subspace  $W$ of dimension $(h-\eps)t-\ell t-1$ which contains all these $\lfloor \frac{h-\eps}{\ell} \rfloor-2$  subspaces. Then
for any subspace $T\in \cT$, there is a hyperplane of $\F_q^{(h-\eps)t}$ containing both $W$ and $T$.
Note that there are $\sbinom{\ell t+1}{\ell t}=\sbinom{\ell t +1}{1}$ hyperplanes containing $W$ and each of them contains at most $\alpha-1$ subspaces from $\cT$. Thus
\begin{align*}
  |\cT| \leq  &\parenv*{\floor*{ \frac{h-\eps}{\ell} }-2} \\
  &+ \sbinom{\ell t+1}{\ell t}_q\parenv*{\alpha-1 - \parenv*{\floor*{ \frac{h-\eps}{\ell} }-2}} \\
  = & \parenv*{\floor*{ \frac{h-\eps}{\ell} }-2} +\parenv*{\alpha- \floor*{ \frac{h-\eps}{\ell} }+1}  \sbinom{\ell t+1}{1}_q.
\end{align*}
\end{IEEEproof}

\begin{theorem}\label{thm-upbound-v2}
Let $\alpha\geq 2$, $h,\ell,t\geq 1$, $\eps\geq 0$, $h-\eps\geq 2\ell$, and let $\cS$ be a collection of  subspaces of $\F_q^{ht}$ such that
\begin{enumerate}
\item[(i)] each subspace has dimension at most $\ell t$; and
\item[(ii)] any subset of $\alpha$ subspaces spans a subspace of dimension at least $(h-\eps)t$.
\end{enumerate}
Then  we have $\alpha \ell  \geq h-\eps$ and
\begin{align*}|\cS| \leq &\sbinom{(\eps+\ell)t}{\eps t}_q \parenv*{ \parenv*{\alpha- \floor*{ \frac{h-\eps}{\ell} }+1}  \frac{q^{\ell t+1}-1}{q-1}-1}\\
                         &+ \floor*{\frac{h-\eps}{\ell}}-1\\
\overset{(*)}{<}& \gamma\parenv*{\alpha- \floor*{ \frac{h-\eps}{\ell} }+1} q^{\ell t (\eps t+1)} {+ \floor*{ \frac{h-\eps}{\ell}}-1}.
\end{align*}
\end{theorem}

\begin{IEEEproof}
Take arbitrarily $\floor*{\frac{h-\eps}{\ell}}-1$ subspaces from $\cS$ and a subspace $W \subset \F_q^{ht}$ of dimension $(h-\eps)t-\ell t$ such that $W$ contains all these  $\floor*{\frac{h-\eps}{\ell}}-1$ subspaces.  Then for any subspace $S \in \cS$ there is a subspace of dimension $(h-\eps)t$ containing both $W$ and $S$.

Let $m:= \sbinom{(\eps+\ell )t}{\eps t}_q$. Then there are $m$ subspaces of dimension $(h-\eps)t$ containing $W$, say $W_1, W_2, \ldots, W_m$. Note that every $\alpha$ subspaces in $W_i \cap \cS$ span the subspace $W_i$. According to Lemma~\ref{lm-full-t2}, we have
\begin{align*}
    |W_i \cap \cS| &\leq \parenv*{\floor*{ \frac{h-\eps}{\ell} }-2}
    +\parenv*{\alpha- \floor*{ \frac{h-\eps}{\ell} }+1}  \sbinom{\ell t +1}{1}_q.
\end{align*}
Hence,
\begin{align*}
|\cS|  \leq& \sum_{i=1}^{m}\parenv*{|W_i\cap \cS|- \parenv*{\floor*{ \frac{h-\eps}{\ell}}-1  }} + \floor*{ \frac{h-\eps}{\ell}}-1 \\
  \leq & \sbinom{(\eps+\ell)t}{\eps t}_q \parenv*{ \parenv*{\alpha- \floor*{ \frac{h-\eps}{\ell} }+1}  \frac{q^{\ell t+1}-1}{q-1}  -1}\\
         &+ \floor*{ \frac{h-\eps}{\ell}}-1.
 \end{align*}
The inequality $(*)$ is derived by \eqref{eq:gauss}.
\end{IEEEproof}
The following corollary rephrases Theorem~\ref{thm-upbound-v2} with network parameters.

\begin{corollary}\label{cor:imupperbound-N}
Let $\alpha\geq 2$, $h,\ell,t\geq 1$, $\eps\geq 0$, and $h-\eps \geq 2\ell$. If $(\eps,\ell)-\mathcal{N}_{h,r,\alpha\ell+\eps}$ has a $(q,t)$-linear solution then
\[
r \leq r_{\max}
< \gamma\theta q^{\ell t (\eps t+1)} +\alpha-\theta,
\]
where $\theta:= \alpha- \floor*{ \frac{h-\eps}{\ell} }+1$ and $\gamma\approx 3.48$.
\end{corollary}
\begin{IEEEproof}
If a $(q,t)$-linear solution exists, then each of the $r$ nodes in the middle layer gets a subspace of dimension $\ell t$ of the source messages space. Since all receivers are able to recover the entire source message space, every $\alpha$-subset of the middle nodes span a subspace of dimension at least $(h-\eps)t$. We then use Theorem~\ref{thm-upbound-v2}.
\end{IEEEproof}

Theorem~\ref{thm-upbound-v2} and Corollary~\ref{cor:imupperbound-N} are valid for all $\alpha\geq 2$. However, we derive a better upper bound for $\alpha = 2$, as shown in the following theorem.

\begin{theorem}\label{thm:imupbound-2}
Let $\alpha=2,h,\ell,t\geq 1$, $\eps\geq 0$, and let $\cS$ be a collection of  subspaces of $\F_q^{ht}$ such that
\begin{enumerate}
\item[(i)] each subspace has dimension at most $\ell t$; and
\item[(ii)] the sum of any two subspaces has dimension at least $(h-\eps)t$.
\end{enumerate}
Then we have
\[|\cS|  \leq \frac{\sbinom{ht}{2\ell t - (h-\eps)t+1}_q}{ \sbinom{\ell t}{2\ell t - (h-\eps)t+1}_q}
 < \gamma \cdot q^{(h-\ell)(2\ell+\eps-h)t^2+(h-\ell)t}.
\]
\end{theorem}

\begin{IEEEproof}
We may assume that each subspace has dimension  $\ell t$. Since the sum of every two subspaces has dimension at least $(h-\eps)t$, then their intersection has dimension at most $2\ell t - (h-\eps)t$. It follows that any subspace of dimension $2\ell t - (h-\eps)t+1$ is contained in at most one subspace of $\cS$. Note that there are $\sbinom{ht}{2\ell t - (h-\eps)t+1}_q$ subspaces of dimension  $2\ell t - (h-\eps)t+1$  and each   subspace of dimension $\ell t$ contains $\sbinom{\ell t}{2\ell t - (h-\eps)t+1}_q$ such spaces.  We have that
$$|\cS| \leq \sbinom{ht}{2\ell t - (h-\eps)t+1}_q \Big/ \sbinom{\ell t}{2\ell t - (h-\eps)t+1}_q.$$
\end{IEEEproof}
The following corollary rephrases Theorem~\ref{thm:imupbound-2} with network parameters.

\begin{corollary}\label{cor:imupperbound-2-Network}
Let $\alpha= 2$, $h,\ell,t\geq 1$, $\eps\geq 0$. If $(\eps,\ell)-\mathcal{N}_{h,r,\alpha\ell+\eps}$ has a $(q,t)$-linear solution then
\[
r \leq r_{\max}
< \gamma \cdot q^{(h-\ell)(2\ell+\eps-h)t^2+(h-\ell)t},
\]
where $\gamma\approx 3.48$.
\end{corollary}
\begin{IEEEproof}
If a $(q,t)$-linear solution exists, then each of the $r$ nodes in the middle layer gets a subspace of dimension $\ell t$ of the source messages space. Since all receivers are able to recover the entire source message space, any two subset of the middle nodes span a subspace of dimension at least $(h-\eps)t$. We then use Theorem~\ref{thm:imupbound-2}.
\end{IEEEproof}

\section{Lower Bounds on the Maximum Number of Middle Layer Nodes}

\label{sec:lb_rv}

We now turn to study a lower bound on $r_{\max}$ {with the parameters $\alpha,\ell,\eps,h$ being fixed}. The main results are summarized in Theorem~\ref{thm:LLL_bound} and Corollary~\ref{cor:EK19_lb}.
In the following, we first give the condition on the coding coefficients under which a linear solution exists.

Let $\x_1,\dots,\x_h\in\mathbb{F}^t_{q}$ denote the $h$ source messages and $\y_1,\dots,\y_N\in\mathbb{F}^{(\eps+\alpha\ell)t}_q$ the messages received by each receiver\footnote{The vector $\y_i$ is the concatenation of all the messages received by the $i$th receiver node.}. Since each middle-layer node receives $\ell$ incoming edges, and has $\ell$ outgoing edges directed at a given receiver, we may assume without loss of generality that this node just forwards its incoming messages. Let us denote the \emph{coding coefficients} used by the source node for the messages transmitted to the $r$ middle nodes by $\A_1,\dots,\A_r \in \Fq^{\ell t \times ht}$. Additionally, we denote the coding coefficients used by the source node for the messages transmitted directly to the receivers by $\B_1,\dots,\B_N\in\F_q^{\eps t\times ht}$.

Each receiver has to solve the following linear system of equations (LSE):
\begin{equation*}
\y_i
=\begin{pmatrix}
  \A_{i_1}\\ \vdots\\ \A_{i_\alpha}\\ \B_i
\end{pmatrix}_{(\eps+\alpha\ell)t\times ht}
\cdot
\begin{pmatrix}
\x_1\\
\vdots\\
\x_{h}
\end{pmatrix}_{ht\times 1},\ \forall i =1,\dots,N=\binom{r}{\alpha},
\end{equation*}
where $\{\A_{i_1},\dots, \A_{i_\alpha}\}\subset \{\A_1,\dots,\A_r \}$.

Any receiver can recover the $h$ source messages $\x_1,\dots,\x_h$ if and only if
\begin{equation}\label{eq:NC_sol}
\rank
\begin{pmatrix}
  \A_{i_1}\\ \vdots\\ \A_{i_\alpha}
\end{pmatrix}_{\alpha \ell t\times ht}
\geq (h-\eps)t,\ \forall i =1,\dots,N.
\end{equation}
Here the {solution} of the $(\eps,\ell)-\mathcal{N}_{h,r,\alpha\ell+\eps}$ network is a set of the coding coefficients $\{\A_1,\dots,\A_r \}$ s.t.~\eqref{eq:NC_sol} holds (where $\B_1,\dots,\B_N$ may be easily determined from the solution).

\subsection{A Lower Bound by the Lov\'asz-Local Lemma}
\begin{lemma}[The Lov\'asz-Local-Lemma~{\cite[Ch.~5]{TheProbMethod}}, \cite{LLLbeck1991}]\label{lem:LLL}
Let $\mathcal{E}_1, \mathcal{E}_2, \hdots,\mathcal{E}_k$ be a sequence of events. Each event occurs with probability at most $p$ and each event is independent of all the other events except for at most $d$ of them.
If $epd\leq 1$ {(where $e\approx 2.71828$ is Euler's number)}, then there is a non-zero probability that none of the events occurs.
\end{lemma}

We choose the matrices $\A_1,\dots,\A_r \in \Fq^{\ell t \times ht}$ independently and uniformly at random.
For $1 \leq i_1 <  \dots<i_\alpha \leq r$, we define the event
\begin{align*}
\mathcal{E}_{i_1,\dots,i_\alpha} := \set*{ (\A_{i_1},\dots, \A_{i_\alpha}) ; \rank \begin{pmatrix}
\A_{i_1} \\ \vdots \\ \A_{i_\alpha}
\end{pmatrix} < (h-\eps)t }.
\end{align*}

Let $p=\Pr(\mathcal{E}_{i_1,\dots,i_\alpha})$ and denote by $d$ the number of other events $\mathcal{E}_{i'_1,\dots,i'_\alpha}$ that are dependent on $\mathcal{E}_{i_1,\dots,i_\alpha}$.

\begin{lemma}\label{lem:upper_bound_on_p}
Let $\alpha\geq 2$, $h,\ell,t\geq 1$, $\eps\geq 0$. Fixing $1 \leq i_1 < \dots < i_\alpha \leq r$, we have
\begin{align*}
\Pr(\mathcal{E}_{i_1,\dots,i_{\alpha}}) \leq 2\gamma \cdot q^{(h-\alpha\ell-\eps)\eps t^2+(h-\alpha\ell-2\eps)t-1},
\end{align*}
where $\gamma\approx3.48$.
\end{lemma}
\begin{IEEEproof}
The number of matrices $\A\in\mathbb{F}_q^{m\times n}$ of rank $s$ is
\begin{equation}
  M(m,n,s):=\prod\limits^{s-1}_{j=0}\frac{(q^m-q^j)(q^n-q^j)}{q^s-q^j}
  \leq \gamma\cdot q^{(m+n)s-s^2} \label{eq:number_matrices_upper_bound}.
\end{equation}
Then,
\begin{align}
  \Pr(\mathcal{E}_{i_1,\dots,i_{\alpha}}) &=\frac{\sum\limits^{(h-\eps)t-1}_{i=0}M(\alpha \ell t,ht,i)}{q^{\alpha \ell h t^2}}\nonumber\\
    & \leq \frac{\sum\limits^{(h-\eps)t-1}_{i=0}\gamma \cdot q^{(h+\alpha\ell)ti-i^2}}{q^{\alpha\ell h t^2}}\label{eq:NM_upper_bound} \\
    & \leq \gamma \cdot \frac{q}{q-1}\cdot q^{\max_i\{(h+\alpha\ell)ti-i^2\} - \alpha\ell h t^2} \label{eq:summation_upper_bound}\\
    & = \gamma \cdot \frac{q}{q-1}\cdot q^{(h+\alpha\ell)ti-i^2|_{i=(h-\eps)t-1} - \alpha\ell h t^2} \label{eq:quadratic_maximum}\\
    & \leq \gamma\cdot 2\cdot q^{(h-\alpha\ell-\eps)\eps t^2+(h-\alpha\ell-2\eps)t-1} \nonumber
\end{align}
where (\ref{eq:NM_upper_bound}) holds due to~\eqref{eq:number_matrices_upper_bound}, (\ref{eq:summation_upper_bound}) follows from a geometric sum, and \eqref{eq:quadratic_maximum} follows by maximizing $(h+\alpha\ell)ti-i^2$.
\end{IEEEproof}

\begin{lemma}\label{lem:upper_bound_on_d}
Let $\alpha\geq 2$, $h,\ell,t\geq 1$, $\eps\geq 0$. Fixing $1 \leq i_1 <\dots < i_\alpha \leq r$, the event $\mathcal{E}_{i_1,\dots,i_\alpha}$ is statistically independent of all the other events $\mathcal{E}_{i_1',\dots,i_\alpha'}$ ($1 \leq i_1' < \dots < i_\alpha' \leq r$), except for at most $\alpha\binom{r-1}{\alpha-1}$
of them.
\end{lemma}

\begin{IEEEproof}
For $1 \leq i_1 < \dots < i_\alpha \leq r$ and $1 \leq i_1' <\dots < i_\alpha' \leq r$, the events $\mathcal{E}_{i_1,\dots,i_\alpha}$ and $\mathcal{E}_{i_1',\dots,i_\alpha'}$ are statistically independent if and only if $\{i_1,\dots,i_\alpha\} \cap \{i_1',\dots,i_\alpha'\} = \emptyset$. Thus, having chosen $1 \leq i_1 < \dots < i_\alpha \leq r$, there are at most
$\alpha\binom{r-1}{\alpha-1}$
ways of choosing an independent event.
\end{IEEEproof}
\begin{remark}
Lemma~\ref{lem:upper_bound_on_d} is a union-bound argument on the number of dependent events. The exact number is $\binom{r}{\alpha}-\binom{r-\alpha}{\alpha}$.
 However the exact expression makes it harder to resolve everything for $r$ later so we use the bound here.
\end{remark}
\begin{theorem}\label{thm:LLL_bound}
  Let $\alpha\geq 2$, $\eps \geq 0$, $\ell,t \geq 1$, and $1\leq h \leq \alpha\ell+\eps$ be fixed integers.
  If
  \begin{equation}\label{eq:r_LLL_lb}
      r \leq \beta \cdot
      q^{\frac{f(t)}{\alpha-1}}
  \end{equation}
where $\beta := \parenv*{\frac{(\alpha-1)!}{2e\gamma\alpha}}^{\frac{1}{\alpha-1}}, \gamma\approx3.48$, {$e\approx 2.71828$ is Euler's number,} and $f(t):=(\alpha\ell+\eps-h)\eps t^2+(\alpha\ell+2\eps-h)t +{1}$, then $(\eps,\ell)-\mathcal{N}_{h,r,\alpha\ell+\eps}$ has a $(q,t)$-linear solution.

Namely, for an $(\eps,\ell)-\mathcal{N}_{h,r,\alpha\ell+\eps}$ that has a $(q,t)$-linear solution, the maximum number of middle nodes satisfies
\[
      r_{\max}\geq \beta \cdot q^{\frac{f(t)}{\alpha-1}}.
\]
\end{theorem}

\begin{IEEEproof}
By the Lov\'asz Local Lemma, it suffices to show that $epd\leq 1$. Noting that $d\leq \alpha\binom{r-1}{\alpha-1}\leq \alpha\cdot \frac{(r-1)^{\alpha-1}}{(\alpha-1)!}$, we shall require
\begin{align*}
    e\cdot2\gamma q^{(h-\alpha\ell-\eps)\eps t^2+(h-\alpha\ell-2\eps)t-1}
    \cdot \alpha\frac{(r-1)^{\alpha-1}}{(\alpha-1)!}\leq 1.
\end{align*}

Namely, if
$ r \leq {\beta}\cdot
q^{\frac{(\alpha\ell+\eps-h)\eps}{\alpha-1}t^2+\frac{\alpha\ell+2\eps-h}{\alpha-1}t+\frac{1}{\alpha-1}}+1$, then $(\eps,\ell)-\mathcal{N}_{h,r,\alpha\ell+\eps}$ has a $(q,t)$-linear solution.
We omit the plus one for simplicity.
\end{IEEEproof}
\begin{remark}
 For any $\alpha\geq 7$,
 \eqref{eq:r_LLL_lb} can be simplified to
 \begin{equation*}
    r \leq
    q^{\frac{f(t)}{\alpha-1}},
 \end{equation*}
 since the prefactor $\beta>1$ for all $\alpha\geq 7$.
\end{remark}

\begin{remark}
 For $t\geq 3$, $\alpha \geq 5$ or $q\geq 4$, it can be seen from numerical analysis that
 $\beta \cdot q^{\frac{\alpha\ell+2\eps-h}{\alpha-1}t +\frac{1}{\alpha-1}} \geq 1$. Thus, \eqref{eq:r_LLL_lb} can be simplified to a looser upper bound
 \[
    r \leq q^{\frac{(\alpha\ell+\eps-h)\eps}{\alpha-1}t^2}.
 \]
 However, omitting the term $\beta \cdot q^{\frac{\alpha\ell+2\eps-h}{\alpha-1}t+\frac{1}{\alpha-1}} $ will cause a loss in estimating the maximum achievable number of middle nodes. Nevertheless, the loss is negligible when $t\to \infty$.
\end{remark}

\subsection{A Lower Bound by $\alpha$-Covering Grassmannian Codes}
\begin{definition}[Covering Grassmannian Codes~\cite{EZjul2019}]
  An $\alpha$-$(n,k,\delta)_q^c$ \emph{covering Grassmannian code} $\mathbb{C}$ is a subset of $\mathcal{G}(n,k)$ such that each subset with $\alpha$ codewords of $\mathbb{C}$ spans a subspace whose dimension is at least $\delta+k$ in $\mathbb{F}_q^n$. Additionally, let ${\cal B}_q(n,k,\delta;\alpha)$ denote the maximum possible size of an $\alpha$-$(n,k,\delta)_q^c$ covering Grassmannian code.
\end{definition}
The following theorem from~\cite{EZjul2019} shows the connection between covering Grassmannian codes and linear network coding solutions.
\begin{theorem}[{\cite[Thm.~4]{EZjul2019}}]\label{thm:cover_scalar_sol}
  The $(\eps,\ell)-\mathcal{N}_{h,r,\alpha\ell+\eps}$ network is solvable with a $(q,t)$-linear solution if and only if there exists an $\alpha$-$(ht,\ell t,ht-\ell t-\eps t)_{q}^c$ code with $r$ codewords.
\end{theorem}

{
For two matrices $A,B\in\F_q^{k\times\ell}$, the rank distance between them is defined to be $d(A,B):=\rank(A-B)$. A linear subspace $\mathbb{C}\subseteq \F_q^{k\times\ell}$ is a linear rank-metric code with parameters $[k\times\ell,K,d]$ if it has dimension $K$, and for any two distinct matrices $C_1,C_2\in\mathbb{C}$, $d(C_1,C_2)\geq d$. It was proved in~\cite{Del78} that
\[ K\leq \min\{ k(\ell-d+1),\ell(k-d+1)\}.\]
Codes attaining this bound with equality are always possible, and are called \emph{maximum rank distance (MRD) codes~\cite{Del78}}.
}

Let $\A$ be a $k\times (n-k)$ matrix, and let $\ve{I}_k$ be a $k\times k$ identity matrix. The matrix $[\ve{I}_k\ \A]$ can be viewed as a generator matrix of a $k$-dimensional subspace of $\Fq^{n}$, and it is called the \emph{lifting} of $\A$.
When all the codewords of an MRD code are lifted to $k$-dimensional subspaces, the result is
called a \emph{lifted MRD code}, denoted by $\mathbb{C}^{\mathrm{MRD}}$.
\begin{theorem}\label{thm:EK_1_ext}
Let $n,k,\delta$ and $\alpha$ be positive integers such that $1\leq \delta \leq k$, $\delta+k\leq n$ and $\alpha\geq 2$. Then
$${\cal B}_q(n,k,\delta;\alpha) \geq (\alpha -1) q^{\max\{k,n-k\}(\min\{k,n-k\}-\delta+1)}.$$
\end{theorem}

\begin{IEEEproof} Let $m=n-k$ and $K=\max\{m,n-m\}(\min\{m,n-m\}-\delta+1)$.
Since $\delta \leq \min\{m,n-m\}$,  an $[m\times (n-m), K, \delta]_q$ MRD code $\mathbb{C}$ exists.
Let $\mathbb{C}^{\mathrm{MRD}}$ be the lifted code of $\mathbb{C}$. Then $\mathbb{C}^{\mathrm{MRD}}$ is a subspace code of $\Fq^n$, which contains $q^K$ $m$-dimensional subspaces as codewords and its minimum subspace distance is $2\delta$~\cite{SKK08}.
Hence, for any two different codewords $C_1,C_2 \in \mathbb{C}^{\mathrm{MRD}}$ we have
$$\dim (C_1 \cap C_2)\leq m-\delta.$$

Now, let $\mathbb{D}=\set*{C^{\perp} ; C \in \mathbb{C}^{\mathrm{MRD}}}$. Take $\alpha -1$ copies of $\mathbb D$ and denote their multiset union as  $\mathbb{D}^{(\alpha)}$. We claim that $\mathbb{D}^{(\alpha)}$ is an $\alpha$-$(n,k,\delta)_q^c$ covering Grassmannian code. For each codeword of $\mathbb{D}^{(\alpha)}$, since it is the dual of a codeword in $\mathbb{C}^{\mathrm{MRD}}$, it has dimension $n-m$, which is $k$. For arbitrarily $\alpha$ codewords $D_1,D_2,\ldots, D_\alpha$ of $\mathbb{D}^{(\alpha)}$, there exist $1\leq i < j\leq \alpha$ such that $D_i \not= D_j$. Let $C_i=D_i^{\perp}$ and $C_j=D_j^\perp$. Then $C_i$ and $C_j$ are two distinct codewords of $\mathbb{C}^{\mathrm{MRD}}$. It follows that
\begin{align*}
    \dim\parenv*{\sum_{\ell=1}^\alpha D_\ell} & \geq \dim \parenv*{D_i +D_j }
    = n - \dim \parenv*{D_i^{\perp} \cap D_j^{\perp}}\\
    & = n - \dim \parenv*{C_i \cap C_j}
    \geq n-m+\delta = k+\delta.
\end{align*}
So far we have shown that $\mathbb{D}^{(\alpha)}$ is an $\alpha$-$(n,k,\delta)_q^c$ covering Grassmannian code. Then the conclusion follows by noting that
\begin{align*}
|\mathbb{D}^{(\alpha)}|= & (\alpha-1)|\mathbb{D}|=(\alpha-1)|\mathbb{C}^{\mathrm{MRD}}| \\
= & (\alpha-1) q^{\max\{k,n-k\}(\min\{k,n-k\}-\delta+1)}.
\end{align*}
\end{IEEEproof}

Corollary~\ref{cor:EK19_lb} below results from the relation between covering Grassmannian codes and network solutions in Theorem~\ref{thm:cover_scalar_sol} and the lower bound on the cardinality of covering Grassmannian codes in Theorem~\ref{thm:EK_1_ext}.

\begin{corollary}\label{cor:EK19_lb}
Let $\alpha\geq 2$, $h,\ell,t\geq 1$, $\eps\geq 0$, $h\leq 2\ell+\eps$.
For an $(\eps,\ell)-\mathcal{N}_{h,r,\alpha\ell+\eps}$ which has a $(q,t)$-linear solution, the maximum number of middle nodes
\[
r_{\max}\geq (\alpha-1)q^{g(t)}
\]
where
 \begin{align*}
     g(t):=&\max\{\ell t,(h-\ell)t\}\\
     &\cdot(\min\{\ell t, (h-\ell)t\}-(h-\ell-\eps)t+1) \\
     =&\begin{cases}\ell\eps t^2 +\ell t & h\leq 2\ell, \\ (h-\ell)(2\ell+\eps-h)t^2+(h-\ell)t & \text{otherwise.} \end{cases}
 \end{align*}
\end{corollary}

\section{Bounds on the Field Size Gap}\label{sec:bound_gap}
In previous sections, we discussed bounds on $r_{\max}$. The main results in this section are the lower and upper bounds on $\gap(\cN)$ in Theorem~\ref{thm:gap_lb} and~\ref{thm:gap_ub} respectively.
To discuss $\gap (\cN)$, we first need the following conditions on the smallest field size $q_s(\cN)$ or $q_v(\cN)$, for which a network $\cN$ is solvable.

\begin{lemma}\label{lem:lb_q}
Let $\alpha\geq 2$, $r,h,\ell,t\geq 1$, $\eps\geq 0$. If $(\eps,\ell)-\mathcal{N}_{h,r,\alpha\ell+\eps}$ has a $(q,t)$-linear solution then
    \begin{align*}
        q^t \geq
        \begin{cases}
        \parenv*{ \frac{r+\theta-\alpha}{\gamma\cdot\theta}}^{\frac{1}{\ell (\eps t+1)}} & h\geq 2\ell+\eps,\\
        \parenv*{ \frac{r}{\gamma(\alpha-1)}}^{\frac{1}{\ell(\eps t+1)}} & \text{otherwise,}
        \end{cases}
    \end{align*}
    where $\theta := \alpha-\floor*{\frac{h-\eps}{\ell}}+1$ and $\gamma\approx 3.48$.
\end{lemma}
\begin{IEEEproof}
    The first case follows from Corollary~\ref{cor:imupperbound-N} that for $h\geq 2\ell+\eps$,
    $q^t\geq \parenv*{ \frac{r+\theta-\alpha}{\gamma\cdot\theta}}^{\frac{1}{\ell (\eps t+1)}}$. The second case is derived from an upper bound on $r$ in~\cite{EZjul2019} (recalled in Corollary~\ref{cor:EZ19_vector}) in a similar manner.
\end{IEEEproof}

\begin{lemma}\label{lem:ub_q}
Let $\alpha\geq 2$, $r,h,\ell,t\geq 1$, $\eps\geq 0$. There exists a $(q,t)$-linear solution to $(\eps,\ell)-\mathcal{N}_{h,r,\alpha\ell+\eps}$ when
\[
    q^t \geq
    \begin{cases}
    \parenv*{\frac{r}{\beta}}^{\frac{(\alpha-1)t}{f(t)}} & h\geq2\ell+\eps \\
    \parenv*{\frac{r}{\alpha-1}}^{\frac{t}{g(t)}}   &\text{otherwise,}
    \end{cases}
    \]
    where $\beta$ and $f(t)$ are defined as in Theorem~\ref{thm:LLL_bound}, and $g(t)$ is defined as in Corollary~\ref{cor:EK19_lb}.
\end{lemma}
\begin{IEEEproof}
    The proof is similar to that in Lemma~\ref{lem:lb_q} and the cases follow from Theorem~\ref{thm:LLL_bound} and Corollary~\ref{cor:EK19_lb} respectively.
\end{IEEEproof}
Lemma~\ref{lem:lb_q} and Lemma~\ref{lem:ub_q} can be seen as the necessary and the sufficient conditions respectively on the pair $(q,t)$ s.t.~a $(q,t)$-linear solution exists.

In the following, we use the lemmas above to derive  bounds on the $\gap(\cN)$ for a given network $\cN$. The bounds are determined only by the network parameters.

\begin{theorem}\label{thm:gap_ub}
 Let $\alpha\geq 2$, $r,h,\ell\geq 1$, $\eps\geq 0$. Then for the $(\eps,\ell)-\mathcal{N}_{h,r,\alpha\ell+\eps}$ network,
  \begin{align*}
      \gap(\cN)\leq
      \begin{cases}
      \frac{\alpha-1}{f(1)}\log_2\parenv*{\frac{r}{\beta}}-A & h\geq 2\ell+\eps\\
      \frac{1}{g(1)}\log_2\parenv*{\frac{r}{\alpha-1}}- B & \text{otherwise},
      \end{cases}
  \end{align*}
where $\theta := \alpha-\floor*{\frac{h-\eps}{\ell}}+1$, $\beta$ and $f(t)$ are defined as in Theorem~\ref{thm:LLL_bound}, $g(t)$ is defined as in Corollary~\ref{cor:EK19_lb}, and we define
\begin{align*}
  A := \min \set*{ \log_2\parenv*{q^t} ; q^t \geq \parenv*{\tfrac{r+\theta-\alpha}{\gamma \theta}}^{\frac{1}{\ell(\varepsilon t + 1)}} }
\end{align*}
and
\begin{align*}
B := \min \set*{ \log_2\parenv*{q^t} ; q^t \geq \parenv*{\tfrac{r}{\gamma (\alpha-1)}}^{\frac{1}{\ell(\varepsilon t + 1)}} }.
\end{align*}
Furthermore, for $t_A := \min \set*{ t ; 2^t \geq \parenv*{\tfrac{r+\theta-\alpha}{\gamma \theta}}^{\frac{1}{\ell(\varepsilon t + 1)}} } > 2$, we have
\begin{align*}
A \geq \min\set*{t_A,\frac{1}{\ell(\varepsilon (t_A-2) + 1)} \log_2\parenv*{\tfrac{r+\theta-\alpha}{\gamma \theta}}} \geq t_A-1,
\end{align*}
and for $t_B := \min \set*{ t ; 2^t \geq \parenv*{\tfrac{r}{\gamma (\alpha-1)}}^{\frac{1}{\ell(\varepsilon t + 1)}} } > 2$, we have
\begin{align*}
B \geq \min\set*{t_B,\frac{1}{\ell(\varepsilon (t_B-2) + 1)} \log_2\parenv*{\tfrac{r+\theta-\alpha}{\gamma \theta}}} \geq t_B-1.
\end{align*}

\end{theorem}

\begin{IEEEproof}
We only prove the bound for the case $h\geq 2\ell+\varepsilon$.
The other case follows analogously.
Lemma~\ref{lem:ub_q} implies that
\[q_s(\cN)\leq \parenv*{\frac{r}{\beta}}^{\frac{\alpha-1}{f(1)}}.\]
By the definition of $q_v(\cN)$ and Lemma~\ref{lem:lb_q}, $q^t = q_v(\cN)$ must fulfill
\[q^t \geq \parenv*{\tfrac{r+\theta-\alpha}{\gamma \theta}}^{\frac{1}{\ell(\varepsilon t + 1)}}.\]
Hence, we get a lower bound on $q_v(\cN)$ by determining the smallest $q^t$ that fulfills this inequality, i.e., $A$.
Note that the left-hand side of the inequality is a strictly monotonically increasing function in $t$ (for a fixed $q$), and the right side is monotonically decreasing in $t$ (among others, this implies that $A$ and $t_A$ are well-defined).

For the lower bound on $A$ for $t_A>2$, consider the case that there is a prime power $q>2$ and a positive integer $t$ with $2^{t_A} \geq q^t \geq
\parenv*{\tfrac{r+\theta-\alpha}{\gamma \theta}}^{\frac{1}{\ell(\varepsilon t + 1)}}$.
Then we have $t \leq t_A-2$ since $q \geq 3$ and $t_A \geq 3$.
Hence,
\begin{align*}
q^t \geq \parenv*{\tfrac{r+\theta-\alpha}{\gamma \theta}}^{\frac{1}{\ell(\varepsilon (t_A-2) + 1)}} \geq \parenv*{\tfrac{r+\theta-\alpha}{\gamma \theta}}^{\frac{1}{\ell(\varepsilon (t_A-1) + 1)}} \geq 2^{t_A-1},
\end{align*}
which proves the claim.
\end{IEEEproof}

\begin{corollary}\label{cor:gap_ub}
  Let $\alpha\geq 2$, $r,h,\ell\geq 1$, $\eps\geq 0$. Then for the $(\eps,\ell)-\mathcal{N}_{h,r,\alpha\ell+\eps}$ network,
  \begin{align*}
    \gap(\cN)\leq
      \begin{cases}
      \frac{\alpha-1}{f(1)}\log_2\parenv*{\frac{r}{\beta}}- \max\set*{t'-1,1} &\!\!\! h\geq 2\ell+\eps\\
      \frac{1}{g(1)}\log_2\parenv*{\frac{r}{\alpha-1}}- \max\set*{t''-1,1} & \text{otherwise},
      \end{cases}
  \end{align*}
  %%%% old equation, was too long for double column
  % \begin{align*}
  %     \gap(\cN)\leq
  %     \begin{cases}
  %     \frac{\alpha-1}{f(1)}\log_2\parenv*{\frac{r}{\beta}}- \max\set*{\sqrt{\frac{1}{\ell\varepsilon}\log_2\parenv*{\frac{r+\theta-\alpha}{\gamma\theta}}+\frac{1}{4 \varepsilon^2}}-\frac{2\varepsilon+1}{2 \varepsilon},1} & h\geq 2\ell+\eps\\
  %     \frac{1}{g(1)}\log_2\parenv*{\frac{r}{\alpha-1}}- \max\set*{\sqrt{\frac{1}{\ell\varepsilon}\log_2\parenv*{\tfrac{r}{\gamma (\alpha-1)}}+\frac{1}{4 \varepsilon^2}}-\frac{2\varepsilon+1}{2 \varepsilon},1} & \text{otherwise},
  %     \end{cases}
  % \end{align*}
  where $\gamma\approx 3.48$, $\theta= \alpha-\floor*{\frac{h-\eps}{\ell}}+1$, $\beta= \parenv*{\frac{(\alpha-1)!}{2e\gamma\alpha}}^{\frac{1}{\alpha-1}}$, $f(t)$ and $g(t)$ are defined as in Theoremm~\ref{thm:LLL_bound} and Corollary~\ref{cor:EK19_lb} respectively, and
  \begin{align*}
    t' &= \sqrt{\frac{1}{\ell\varepsilon}\log_2\parenv*{\frac{r+\theta-\alpha}{\gamma\theta}}+\frac{1}{4 \varepsilon^2}}-\frac{1}{2 \varepsilon},\\
  t''&=\sqrt{\frac{1}{\ell\varepsilon}\log_2\parenv*{\tfrac{r}{\gamma (\alpha-1)}}+\frac{1}{4 \varepsilon^2}}-\frac{1}{2 \varepsilon}.
  \end{align*}
  In particular, if all parameters are constants except for $r\to\infty$, then $\gap(\cN)\in O(\log r)$.
\end{corollary}
\begin{IEEEproof}
We only prove the bound for the case $h\geq 2\ell+\varepsilon$.
The other case follows analogously.
We determine $t_A$ as defined in Theorem~\ref{thm:gap_ub}.
Note that $2^t$ is strictly monotonically increasing in $t$ and $\parenv*{\tfrac{r+\theta-\alpha}{\gamma \theta}}^{\frac{1}{\ell(\varepsilon t + 1)}}$ is strictly monotonically decreasing.
Hence, we have $t_A = \lceil t' \rceil$, where $t'$ is the unique (positive) solution of
\[2^{t'} = \parenv*{\tfrac{r+\theta-\alpha}{\gamma \theta}}^{\frac{1}{\ell(\varepsilon t' + 1)}}.\]
By rewriting this equation into a quadratic equation in $t'$, we obtain the following positive solution:
\begin{equation*}
t' = \sqrt{\frac{1}{\ell\varepsilon}\log_2\parenv*{\frac{r+\theta-\alpha}{\gamma\theta}}+\frac{1}{4 \varepsilon^2}}-\frac{1}{2 \varepsilon}.
\end{equation*}
Using the bound $A \geq t_A-1$ for $t_A >2$ (Theorem~\ref{thm:gap_ub}) and the trivial bound $A \geq 1$ otherwise, the claim follows.
The asymptotic statement is an immediate consequence.
\end{IEEEproof}

\begin{theorem}
\label{thm:gap_lb}
  Let $\alpha\geq 2$, $r,h,\ell\geq 1$, $\eps\geq 0$. Then for the $(\eps,\ell)-\mathcal{N}_{h,r,\alpha\ell+\eps}$ network,
  \begin{align*}
      \gap(\cN)\geq
      \begin{cases}
      \frac{1}{\ell(\eps+1)}\log_2\parenv*{\frac{r+\theta-\alpha}{\gamma\theta}}-t_{\Delta} & h\geq 2\ell+\eps\\
      \frac{1}{\ell(\eps+1)}\log_2\parenv*{\frac{r}{\gamma(\alpha-1)}}-t_{\star} & \text{otherwise},
      \end{cases}
  \end{align*}
  where {$\gamma\approx 3.48$, $\theta = \alpha-\floor*{\frac{h-\eps}{\ell}}+1$,} $t_{\Delta}$ is the smallest positive integer s.t.~$2^{\frac{f(t_{\Delta})}{\alpha-1}}\geq \frac{r}{\beta}$ and $t_{\star}$ is the smallest positive integer s.t.~$2^{g(t_{\star})}\geq \frac{r}{\alpha-1}$.
  Here, $\beta$ and $f(t)$ are defined as in Theorem~\ref{thm:LLL_bound}, and $g(t)$ is defined as in Corollary~\ref{cor:EK19_lb}.
\end{theorem}
\begin{IEEEproof}
 Let us only consider the first case $h\geq 2\ell+\eps$. The other case can be proved in the same manner. According to Lemma~\ref{lem:lb_q}, we have the lower bound on the smallest field size of a scalar solution,
 \[q_s(\cN)\geq \parenv*{ \frac{r+\theta-\alpha}{\gamma\cdot\theta}}^{\frac{1}{\ell (\eps +1)}}.\]
For vector solutions, according to Lemma~\ref{lem:ub_q}, we want to find $(q,t)$ s.t.~$q^{\frac{f(t)}{\alpha-1}}\geq \frac{r}{\beta}.$
Since $t_{\Delta}$ is the smallest positive integer $t$ s.t.~$2^{\frac{f(t)}{\alpha-1}}\geq \frac{r}{\beta}$, it is guaranteed that a $(2,t_{\Delta})$-linear solution exists.
Therefore, $q_v(\cN)$ (the smallest value of $q^t$) should be at most
$q_v(\cN)\leq 2^{t_{\Delta}}$.
The lower bound then follows directly from the definition of $\gap(\cN)$.
\end{IEEEproof}

By carefully bounding $t_{\star}$ and $t_{\Delta}$, the following is obtained:

\begin{corollary}
\label{cor:gap}
Let $\alpha\geq 2$, $r,h,\ell,\eps\geq 1$. {Assume that $h\not= \alpha \ell+\varepsilon$.} Then for the $(\eps,\ell)-\mathcal{N}_{h,r,\alpha\ell+\eps}$ network,
\[
\gap(\cN) \geq
\begin{cases}
\frac{\log_2\parenv*{\frac{r+\theta-\alpha}{\gamma \theta}}}{\ell(\eps+1)} - \sqrt{\frac{(\alpha-1)\log_2(\frac{r}{\beta})}{(\alpha\ell+\eps-h)\eps}}  & h\geq 2\ell+\eps\\
\frac{\log_2\parenv*{\frac{r}{\alpha-1}}-2}{\ell(\eps+1)}-\sqrt{\frac{\log_2(\frac{r}{\alpha-1})}{\ell\eps}}& \text{otherwise,}
\end{cases}
\]
{where $\gamma\approx 3.48$, $\theta= \alpha-\floor*{\frac{h-\eps}{\ell}}+1$, $\beta= \parenv*{\frac{(\alpha-1)!}{2e\gamma\alpha}}^{\frac{1}{\alpha-1}}$, $f(t)$ and $g(t)$ are defined as in Theoremm~\ref{thm:LLL_bound} and Corollary~\ref{cor:EK19_lb} respectively.}
In particular, if all parameters are constants except for $r\to\infty$, then $\gap(\cN) \in \Omega(\log r)$.
\end{corollary}
\begin{IEEEproof}
When $h \geq 2\ell+\eps$, noting that $\alpha \ell+2\eps-h>0$ {and $h\not= \alpha \ell+\varepsilon$}, we may choose
$$t=\parenv*{\frac{(\alpha-1)\log_2(\frac{r}{\beta})}{(\alpha\ell+\eps-h)\eps} }^{1/2}$$
such that $2^{f(t)}\geq 2^{(\alpha\ell+\eps-h)\eps t^2}=(\frac{r}{\beta})^{\alpha-1}.$
Then we have that
\begin{align*}
\gap(\cN)\geq &
\frac{\log_2\parenv*{\frac{r+\theta-\alpha}{\gamma \theta}}}{\ell(\eps+1)} - \parenv*{\frac{(\alpha-1)\log_2(\frac{r}{\beta})}{(\alpha\ell+\eps-h)\eps} }^{1/2}
  \\ \geq & \frac{\log_2 (r+\theta-\alpha)-\log_2\theta-2}{\ell(\eps+1)} \\
  &-\parenv*{\frac{\log_2 r-\log_2 \beta}{ (\ell-\frac{h-\ell-\eps}{\alpha-1})\eps}}^{1/2}
\end{align*}
Recall that $\beta$ and $\theta$ are determined by $\alpha, h, \eps,$ and $\ell$. Thus if $\alpha, h, \eps,$ and $\ell$ are fixed,  $\gap(\cN)=\Omega(\log r)$.

When $h< 2\ell+\eps$, we may choose $$t=\parenv*{\frac{\log_2(\frac{r}{\alpha-1})}{\ell\eps}}^{1/2}$$
such that $2^{g(t)}\geq 2^{\ell \eps t^2}=\frac{r}{\alpha-1}.$ It follows that
\begin{align*}
\gap(\cN)&\geq \frac{\log_2\parenv*{\frac{r}{\gamma(\alpha-1)}}}{\ell(\eps+1)}-\parenv*{\frac{\log_2(\frac{r}{\alpha-1})}{\ell\eps}}^{1/2}\\
& \geq \frac{\log_2\parenv*{\frac{r}{\alpha-1}}-2}{\ell(\eps+1)}-\parenv*{\frac{\log_2(\frac{r}{\alpha-1})}{\ell\eps}}^{1/2}
\end{align*}
This shows that $\gap(\cN) \in \Omega(\log r)$.
\end{IEEEproof}

Corollaries~\ref{cor:gap_ub} and \ref{cor:gap} show that for fixed network parameters except for $r$, the gap size grows as
\begin{equation*}
\gap(\cN) = \Theta(\log r) \quad (r \to \infty).
\end{equation*}

\begin{example}
  We illustrate the proof of Theorem~\ref{thm:gap_ub} and Theorem~\ref{thm:gap_lb} by two network examples with $r=8\times 10^5$ in Figure~\ref{fig:q_bound} and $r=8\times 10^6$ in Figure~\ref{fig:q_bound2}. Note that the curves in the figures are not bounds on the gap size. They are the necessary (blue curve) and the sufficient (green curve) condition on $q^t$ such that a $(q,t)$-linear solution exists. Namely, there is no $(q,t)$-linear solution in the region below the blue curve and there must be a $(q,t)$-linear solution in the region above the green curve. Thus the minimum gap of the network $(2,1)-\cN_{12,r,20}$ is determined by the difference between the necessary condition with $t=1$ and the minimum $2^t$ that is in the region above the sufficient condition. Similarly, the maximum gap of the network is determined by the difference between the sufficient condition with $t=1$ and the minimum $2^t$ that is in the region above the necessary condition.

  By comparing the two plots it can be seen that the gap increases as the number of middle node in the network increases.
\begin{figure*}[t!]
    \centering
    \minipage{0.49\textwidth}
    \definecolor{darkgreen}{rgb}{0,0.7,0}
\begin{tikzpicture}
\pgfplotsset{compat = 1.3}
\begin{axis}[
	legend style={nodes={scale=0.7, transform shape}},
	width = \linewidth,
	height = \linewidth,
	title = {$h=12$, $\varepsilon=2$, $\ell=1$, $r=800000$, $\alpha=20$},
	xlabel = {{$t$}},
	xmin = 1,
	xmax = 20,
	ymin = 1,
	ymax = 5000,
	legend pos = north east,
	legend cell align=left,
	ymode=log,
	grid=both,
	ytick={1,10,100,1000,10000},
	yminorticks=true]

\addplot [solid, color=blue, thick, mark=*] table[row sep=\\] {
1.000000 27.544680 \\
2.000000 7.311772 \\
3.000000 4.141513 \\
4.000000 3.020039 \\
5.000000 2.470231 \\
6.000000 2.149396 \\
7.000000 1.940920 \\
8.000000 1.795247 \\
9.000000 1.688008 \\
10.000000 1.605904 \\
11.000000 1.541101 \\
12.000000 1.488691 \\
13.000000 1.445454 \\
14.000000 1.409190 \\
15.000000 1.378347 \\
16.000000 1.351800 \\
17.000000 1.328715 \\
18.000000 1.308458 \\
19.000000 1.290542 \\
20.000000 1.274584 \\
};
\addlegendentry{{$\left(\frac{r+\theta-\alpha}{\gamma\cdot\theta}\right)^{\frac{1}{\ell (\varepsilon t+1)}}$ (Lemma~\ref{lem:lb_q})}};

\addplot [solid, color=darkgreen, thick, mark=diamond] table[row sep=\\] {
1.000000 910.202123 \\
2.000000 72.448517 \\
3.000000 22.388315 \\
4.000000 11.443355 \\
5.000000 7.418817 \\
6.000000 5.480896 \\
7.000000 4.383703 \\
8.000000 3.692574 \\
9.000000 3.223351 \\
10.000000 2.886711 \\
11.000000 2.634796 \\
12.000000 2.439934 \\
13.000000 2.285133 \\
14.000000 2.159441 \\
15.000000 2.055507 \\
16.000000 1.968230 \\
17.000000 1.893969 \\
18.000000 1.830058 \\
19.000000 1.774507 \\
20.000000 1.725796 \\
};
\addlegendentry{{$\left(\frac{r}{\beta}\right)^{\frac{(\alpha-1)t}{f(t)}}$ (Lemma~\ref{lem:ub_q})}};

\addplot [solid, color=orange, thick, mark=star] table[row sep=\\] {
1.000000 2.000000 \\
2.000000 4.000000 \\
3.000000 8.000000 \\
4.000000 16.000000 \\
5.000000 32.000000 \\
6.000000 64.000000 \\
7.000000 128.000000 \\
8.000000 256.000000 \\
9.000000 512.000000 \\
10.000000 1024.000000 \\
11.000000 2048.000000 \\
12.000000 4096.000000 \\
13.000000 8192.000000 \\
14.000000 16384.000000 \\
15.000000 32768.000000 \\
16.000000 65536.000000 \\
17.000000 131072.000000 \\
18.000000 262144.000000 \\
19.000000 524288.000000 \\
20.000000 1048576.000000 \\
};
\addlegendentry{{$2^t$}};

\addplot [dashed, color=black, thick, forget plot] table[row sep=\\] {
1.000000 910.202123 \\
5.000000 910.202123 \\
};
\addlegendentry{{$2^t$}};

\addplot [dashed, color=black, thick, forget plot] table[row sep=\\] {
1.000000 8.000000 \\
5.000000 8.000000 \\
};
\addlegendentry{{$2^t$}};

\draw[<->,>=latex, thick] (axis cs: 4.000000, 910.202123) -- node[right] {max $\gap(\cN) \approx 6.83$ bits} (axis cs: 4.000000, 8.000000);

\addplot [dashed, color=black, thick, forget plot] table[row sep=\\] {
1.000000 27.544680 \\
8.000000 27.544680 \\
};
\addlegendentry{{$2^t$}};

\addplot [dashed, color=black, thick, forget plot] table[row sep=\\] {
1.000000 16.000000 \\
8.000000 16.000000 \\
};
\addlegendentry{{$2^t$}};

\draw[<->,>=latex, thick] (axis cs: 7.000000, 27.544680) -- node[right] {min $\gap(\cN) \approx 0.78$ bits} (axis cs: 7.000000, 16.000000);

\end{axis}
\end{tikzpicture}
    \caption{An illustration of proofs of Theorem~\ref{thm:gap_ub} and Theorem~\ref{thm:gap_lb} for the network $(2,1)-\cN_{12,8e5,20}$.
    }
    \label{fig:q_bound}
\endminipage\hfill
\minipage{0.49\textwidth}
    \definecolor{darkgreen}{rgb}{0,0.7,0}
\begin{tikzpicture}
\pgfplotsset{compat = 1.3}
\begin{axis}[
	legend style={nodes={scale=0.7, transform shape}},
	width = \linewidth,
	height = \linewidth,
	title = {$h=12$, $\varepsilon=2$, $\ell=1$, $r=8000000$, $\alpha=20$},
	xlabel = {{$t$}},
	xmin = 1,
	xmax = 20,
	ymin = 1,
	ymax = 5000,
	legend pos = north east,
	legend cell align=left,
	ymode=log,
	grid=both,
	ytick={1,10,100,1000,10000},
	yminorticks=true]

\addplot [solid, color=blue, thick, mark=*] table[row sep=\\] {
1.000000 59.343413 \\
2.000000 11.588401 \\
3.000000 5.754622 \\
4.000000 3.900535 \\
5.000000 3.045419 \\
6.000000 2.565901 \\
7.000000 2.262948 \\
8.000000 2.055645 \\
9.000000 1.905488 \\
10.000000 1.792004 \\
11.000000 1.703372 \\
12.000000 1.632318 \\
13.000000 1.574133 \\
14.000000 1.525641 \\
15.000000 1.484625 \\
16.000000 1.449492 \\
17.000000 1.419068 \\
18.000000 1.392474 \\
19.000000 1.369031 \\
20.000000 1.348214 \\
};
\addlegendentry{{$\left(\frac{r+\theta-\alpha}{\gamma\cdot\theta}\right)^{\frac{1}{\ell (\varepsilon t+1)}}$ (Lemma~\ref{lem:lb_q})}};

\addplot [solid, color=darkgreen, thick, mark=diamond] table[row sep=\\] {
1.000000 3426.852562 \\
2.000000 166.699487 \\
3.000000 40.991539 \\
4.000000 18.387193 \\
5.000000 10.956583 \\
6.000000 7.631492 \\
7.000000 5.844183 \\
8.000000 4.761175 \\
9.000000 4.047704 \\
10.000000 3.548001 \\
11.000000 3.181350 \\
12.000000 2.902351 \\
13.000000 2.683767 \\
14.000000 2.508383 \\
15.000000 2.364848 \\
16.000000 2.245401 \\
17.000000 2.144574 \\
18.000000 2.058413 \\
19.000000 1.983995 \\
20.000000 1.919113 \\
};
\addlegendentry{{$\left(\frac{r}{\beta}\right)^{\frac{(\alpha-1)t}{f(t)}}$ (Lemma~\ref{lem:ub_q})}};

\addplot [solid, color=orange, thick, mark=star] table[row sep=\\] {
1.000000 2.000000 \\
2.000000 4.000000 \\
3.000000 8.000000 \\
4.000000 16.000000 \\
5.000000 32.000000 \\
6.000000 64.000000 \\
7.000000 128.000000 \\
8.000000 256.000000 \\
9.000000 512.000000 \\
10.000000 1024.000000 \\
11.000000 2048.000000 \\
12.000000 4096.000000 \\
13.000000 8192.000000 \\
14.000000 16384.000000 \\
15.000000 32768.000000 \\
16.000000 65536.000000 \\
17.000000 131072.000000 \\
18.000000 262144.000000 \\
19.000000 524288.000000 \\
20.000000 1048576.000000 \\
};
\addlegendentry{{$2^t$}};

\addplot [dashed, color=black, thick, forget plot] table[row sep=\\] {
1.000000 3426.852562 \\
5.000000 3426.852562 \\
};
\addlegendentry{{$2^t$}};

\addplot [dashed, color=black, thick, forget plot] table[row sep=\\] {
1.000000 8.000000 \\
5.000000 8.000000 \\
};
\addlegendentry{{$2^t$}};

\draw[<->,>=latex, thick] (axis cs: 4.000000, 3426.852562) -- node[right] {max $\gap(\cN) \approx 8.74$ bits} (axis cs: 4.000000, 8.000000);

\addplot [dashed, color=black, thick, forget plot] table[row sep=\\] {
1.000000 59.343413 \\
8.000000 59.343413 \\
};
\addlegendentry{{$2^t$}};

\addplot [dashed, color=black, thick, forget plot] table[row sep=\\] {
1.000000 32.000000 \\
8.000000 32.000000 \\
};
\addlegendentry{{$2^t$}};

\draw[<->,>=latex, thick] (axis cs: 7.000000, 59.343413) -- node[right] {min $\gap(\cN) \approx 0.89$ bits} (axis cs: 7.000000, 32.000000);

\end{axis}
\end{tikzpicture}
    \caption{An illustration of proofs of Theorem~\ref{thm:gap_ub} and Theorem~\ref{thm:gap_lb} for the network $(2,1)-\cN_{12,8e6,20}$.
    }
    \label{fig:q_bound2}
    \endminipage
\end{figure*}
\end{example}

\section{Discussion}\label{sec:discussion}
In this section we will compare our upper and lower bound on $r_{\max}$ with previous known bounds.
\subsection{Other Upper Bound on $r_{\max}$}

In the following we recall the result from~\cite[Corollary 3]{EZjul2019} and compare it with our upper bound in Corollary~\ref{cor:imupperbound-2-Network}.

\begin{theorem}[{\cite[Corollary~3]{EZjul2019}}]\label{thm:codesize}
  If $n$, $k$, $\delta$, and $\alpha$, are positive integers such that $1<k<n,\ 1\leq\delta\leq n-k$ and $2\leq\alpha\leq\quadbinom{k+\delta-1}{k}_{q}+1$, then for an $\alpha-(n,k,\delta)_q^c$ {covering Grassmannian code} $\mathbb{C}$, we have that
$$|\mathbb{C}|\leq\floor*{(\alpha-1)\frac{\quadbinom{n}{\delta+k-1}_{q}}{{\quadbinom{n-k}{\delta-1}_{q}}}}.$$
\end{theorem}

By combining Theorem~\ref{thm:codesize} and Theorem~\ref{thm:cover_scalar_sol}, the following corollary can be directly derived.
\begin{corollary}\label{cor:EZ19_vector}
  If the $(\eps,\ell)-\mathcal{N}_{h,r,\alpha\ell+\eps}$ network has a $(q,t)$-linear solution then
  \begin{align*}
    r\leq r_{\max}&\leq\floor*{(\alpha-1)\frac{\quadbinom{ht}{ht-\eps t-1}_{q}}{{\quadbinom{ht-\ell t}{ht -\ell t-\eps t -1}_{q}}}}\\
    & < (\alpha-1)\frac{\gamma q^{(\eps t+1)(ht -\eps t-1)}}{q^{(\eps t+1)(ht -\ell t -\eps t-1)}}\\
    & =  \gamma (\alpha-1)q^{\ell t(\eps t+1)},
  \end{align*}
  with $1<\ell t<h t$, $0\leq \eps\leq h-\ell-\frac{1}{t}$, $2\leq \alpha\leq\quadbinom{ht-\eps t-1}{\ell t}_{q}+1$.\\
\end{corollary}
\subsection{Comparison Between Upper Bounds}
In the following, we first show that for some parameters, the upper bound in Corollary~\ref{cor:imupperbound-N} could be better than that in Corollary~\ref{cor:EZ19_vector}.
Denote
$$\theta := \parenv*{\alpha- \floor*{ \frac{h-\eps}{\ell} }+1}.$$
The upper bound in Corollary~\ref{cor:imupperbound-N} and Corollary~\ref{cor:EZ19_vector} can be respectively written as
$$U_A := \sbinom{(\eps+\ell) t}{\eps t}_q \parenv*{\theta \cdot \frac{q^{\ell t+1}-1}{q-1}  -1   }+ \alpha -\theta$$
 and
$$U_B:= (\alpha-1)\frac{\quadbinom{ht}{ht-\eps t-1}_{q}}{{\quadbinom{ht-\ell t}{ht -\ell t-\eps t -1}_{q}}}=(\alpha-1)q^{\ell t(\eps t +1)}\prod\limits^{\eps t}_{i=0}\frac{q^{ht-i}-1}{q^{ht-i}-q^{\ell t}}.$$

\begin{lemma}\label{lem:compare-2}
 Let $h \geq 2\ell+\eps$ and $2\leq \alpha \leq \sbinom{ht-\eps t-1}{\ell t}_q+1$. Assume $\sbinom{\eps+\ell)t}{\eps t}_q\leq \alpha$,
then
\[
  \log_q U_A - \log_q U_B < \log_q\frac{2\theta\alpha}{\alpha-1}- \ell \eps t^2.
\]
Particularly, if $$\frac{2\theta \alpha}{\alpha-1}\leq q^{\ell\eps t^2},$$
then
$$U_A<U_B.$$
That is, the upper bound in Corollary~\ref{cor:imupperbound-N} is better than that in  Corollary~\ref{cor:EZ19_vector}.
 \end{lemma}

\begin{IEEEproof}
Under the assumption $\sbinom{(\eps+\ell)t }{\eps t}_q\leq {\alpha}$, we have
\begingroup
\allowdisplaybreaks
\begin{align*}
\log_q U_A & \leq \log_q \parenv*{\alpha \parenv*{ \theta\cdot \frac{q^{\ell t+1}-1}{q-1}  -1   }+ \alpha-\theta } \\
&=\log_q( \alpha \theta \cdot\frac{q^{\ell t+1}-1}{q-1} -\alpha+\alpha-\theta)\\
&=\log_q \theta+\log_q(\alpha \cdot \frac{q^{\ell t+1}-1}{q-1} -1)\\
&<\log_q \theta+\log_q(\alpha \cdot \frac{q^{\ell t+1}-1}{q-1})\\
&\overset{(*)}{<}\log_q \theta +\log_q \alpha + \log_q(2\cdot q^{\ell t})\\
& = \log_q \theta +\log_q \alpha +  \ell t + \log_q 2.
\end{align*}
\endgroup
The inequality $(*)$ is because $\frac{q^{\ell t+1}-1}{q-1}=\sum\limits^{\ell t}_{i=0}q^i<2\cdot q^{\ell t}$.
By the bounds on the $q$-binomial coefficient,
\begin{align*}
  \log_q U_B >  \log(\alpha-1) + \ell t (\eps t+1),
\end{align*}
we have that
\begin{align*}
  \log_q U_A - \log_q U_B < \log_q\frac{2\theta\alpha}{\alpha-1}- \ell \eps t^2.
\end{align*}
Together with the assumption $\frac{2\theta \alpha}{\alpha-1}\leq q^{\ell\eps t^2}$, the conclusion follows.
 \end{IEEEproof}

 \begin{lemma}\label{lem:compare}Let $h \geq 2\ell+\eps$ and $2\leq\alpha \leq \sbinom{ht-\eps t-1}{\ell t}_q+1$.
Assume $\sbinom{(\eps+\ell)t}{\eps t}_q\geq \alpha$. If
$h\geq  2\eps$, then
$$\frac{U_A}{U_B}\leq \frac{8\theta}{\alpha-1}.$$
So, when
$${8\theta} < {\alpha-1},$$
we have that
$${U_A}<{U_B}.$$
 That is, the upper bound in Corollary~\ref{cor:imupperbound-N} is better than that in  Corollary~\ref{cor:EZ19_vector}.
\end{lemma}
\begin{IEEEproof}
  Since $\quadbinom{\eps+\ell)t}{\eps t}_q\geq \alpha$,  we have that
  $$ U_A \leq  \theta\cdot \sbinom{(\eps+\ell) t}{\eps t}_q \frac{q^{\ell t+1}-1}{q-1}.$$
  Then
\begin{align*}
  \frac{U_A}{U_B}  \leq&  \frac{\theta}{\alpha-1}\cdot \frac{q^{\ell t+1}-1}{q-1}\sbinom{(\eps+\ell)t}{\eps t}_q\\
  &\cdot\sbinom{(h-\ell)t}{(h-\ell-\eps)t-1}_q\sbinom{ht}{(h-\eps)t-1}_q^{-1}  \\
  = &\frac{\theta}{\alpha-1}\cdot\frac{q^{\ell t+1}-1}{q-1}\sbinom{(\eps+\ell)t}{\eps t}_q\sbinom{(h-\ell)t}{\eps t+1}_q\sbinom{ht}{\eps t+1}_q^{-1}  \\
  = &\frac{\theta}{\alpha-1}\cdot \frac{q^{\ell t+1}-1}{q-1}\cdot \frac{(q^{(\eps+\ell)t}-1)\cdots (q^{\ell t+1}-1)}{(q^{\eps t}-1)\cdots(q-1)}\\
  &\cdot \frac{(q^{(h-\ell)t}-1)\cdots(q^{(h-\ell-\eps)t}-1)}{(q^{ht}-1)\cdots (q^{(h-\eps)t}-1)}\\
  <& \frac{\theta}{\alpha-1}\cdot \frac{q^{\ell t+1}}{q-1}\cdot \frac{q^{(\eps+\ell)t}\cdots q^{\ell t+1}}{(q^{\eps t}-1)\cdots(q-1)}\\
  &\cdot \frac{q^{(h-\ell)t}\cdots q^{(h-\ell-\eps)t}}{(q^{ht}-1)\cdots (q^{(h-\eps)t}-1)}\\
  =&\frac{\theta}{\alpha-1}\cdot \frac{q}{q-1}\cdot
  \prod_{i=1}^{\eps t}  \parenv*{1-\frac{1}{q^i}}^{-1}  \cdot\prod_{i=h t -\eps t}^{h t}  \parenv*{1-\frac{1}{q^i}}^{-1}\\
  \leq& \frac{\theta}{\alpha-1}\cdot\parenv*{1+\frac{1}{q-1}}
  \prod_{i=1}^{h t}  \parenv*{1-\frac{1}{q^i}}^{-1}\!\textrm{(assume $2\eps\leq h$)}\\
  <&  \frac{8\cdot \theta }{\alpha-1},
\end{align*}
and the conclusion follows.
\end{IEEEproof}

Now,  we compare the upper bound  in Corollary~\ref{cor:imupperbound-2-Network} with that in Corollary~\ref{cor:EZ19_vector} for $\alpha=2$.
\begin{lemma}\label{lem:compare-alpha2}
Denote $U_C:= \gamma q^{(h-\ell)(2\ell+\eps-h)t^2+(h-\ell)t}$
and $U_D:= \gamma q^{\ell t(\eps t+1)}$. Then
\[
    \log_q U_C -\log_q U_D  =   [(h-\ell)(2\ell+\eps-h)-\eps \ell]t^2 +(h-2\ell)t.
\]
Particularly, if one of the following three conditions is satisfied,
\begin{itemize}
    \item $\eps t +1<\ell t$, and either $h>2\ell$ or $h<\ell+\eps+\frac{1}{t}$;
    \item $\eps t +1> \ell t$, and either $h>\ell+\eps+\frac{1}{t}$ or $h<2\ell$;
    \item $\eps t +1=\ell t$ and $h\not=2\ell$,
\end{itemize}
then
\begin{align*}
   \log_q U_C -\log_q U_D   <  0,
\end{align*}
and the upper bound  in Corollary~\ref{cor:imupperbound-2-Network} is better than the upper bound in Corollary~\ref{cor:EZ19_vector} for $\alpha=2$.
\end{lemma}

\begin{IEEEproof}
Denote $C=(h-\ell)(2\ell+\eps-h)t+(h-\eps)$ and $D=\ell(\eps t+1)$. Then $\log_q U_C -\log_q U_D=Ct-Dt$. So it suffices to show that $C < D$.
Note that $C=-th^2 +{3\ell +\eps }t h + h + \cdots$ is a quadratic function in $h$ which is symmetric about $h=\frac{(3\ell+\eps)t+1}{2t}$. We proceed in three cases, according to the position of the axis of symmetry.

\begin{enumerate}
    \item If $\eps t+1 < \ell t$, then $\frac{(3\ell+\eps)t+1}{2t}< 2\ell$, i.e., the axis of symmetry is on the left of $h=2\ell$. In this case, $C$ is decreasing when $h \geq 2\ell$. It follows that $C < D$ for $h> 2\ell$ as $C=D$ when $h=2\ell$.
    Furthermore, according to the symmetry, $C < D$ also holds for $h<\ell +\eps +\frac{1}{t}$.
    \item If $\eps t+1 > \ell t$,  then $\frac{(3\ell+\eps)t+1}{2t}> 2\ell$. Using the same argument, we can see that $C < D$ holds for $h< 2\ell$ and $h>\ell+\eps +\frac{1}{t}$.
    \item If $\eps t+1=\ell t$,  then $\frac{(3\ell+\eps)t+1}{2t}= 2\ell$. The maximal value of $C-D$ is taken at $h=2\ell$, which is $0$. So $C<D$ for all $h\not=2\ell$.
\end{enumerate}
\end{IEEEproof}

The following example shows that, in some cases, the upper bound in Corollary~\ref{cor:imupperbound-2-Network} matches a lower bound from~\cite{EKOOfeb2020} within a factor of $\gamma\approx 3.48$.
\begin{example}
Let $\alpha=2$, $\eps=\ell$, and $h=2\ell+1$.
A lower bound from \cite{EKOOfeb2020} is
$$q^{(\ell^2-1)t^2+(\ell+1) t} \leq r.$$
For the upper bound,
Corollary~\ref{cor:imupperbound-2-Network} shows that
$$r \leq \gamma q^{(\ell^2-1)t^2+(\ell+1)t},$$
agreeing with the lower bound up to a factor of $\gamma$. In contrast, Corollary~\ref{cor:EZ19_vector} shows that
$$r \leq \gamma q^{\ell^2 t^2+\ell t},$$
which differs from the lower bound by a factor of $\gamma q^{t^2-t}$.
\end{example}
\begin{corollary}
  If $n$, $k$, $\delta$, and $\alpha$, are positive integers such that $1<k<n$, $1\leq\delta\leq \min\{ n-k, k\}$, and $2\leq\alpha\leq\quadbinom{k+\delta-1}{k}_{q}+1$, then an upper bound on the size of an $\alpha-(n,k,\delta)_q^c$ code $\mathbb{C}$ is that
  \begin{align*} |{\mathbb{C}}| \leq & \quadbinom{n-\delta}{k}_q \parenv*{ \parenv*{\alpha- \floor*{ \frac{\delta-k}{k} }+1}  \frac{q^{k+1}-1}{q-1}  -1   }\\
    &+ \floor*{ \frac{\delta-k}{k}}-1.
\end{align*}
\end{corollary}
\begin{IEEEproof}
Note that an $\alpha-(n,k,\delta)_q^c$ code $\mathbb{C}$ exists if and only if the $(\eps,\ell)-\mathcal{N}_{h,r,\alpha\ell+\eps}$ network is solvable with linear scalar solutions, where $h=n$, $r=|\mathbb{C}|$, $\ell = k$ and $\eps=h-\ell-\delta=n-k-\delta$. Then the conclusion follows from Corollary~\ref{cor:imupperbound-N} by setting $t=1$.
\end{IEEEproof}

\subsection{Other Lower Bounds on $r_{\max}$}
Let ${\cal B}_q(n,k,\delta;\alpha)$ denote the maximum possible size of an $\alpha$-$(n,k,\delta)_q^c$ covering Grassmannian code. Etzion \emph{et al.} proposed the following lower bounds on  ${\cal B}_q(n,k,\delta;\alpha)$ for $\delta \leq k$ in~\cite{EKOOfeb2020}.

\begin{theorem}[{\cite[Theorem 21]{EKOOfeb2020}}]\label{thm:EK_lb}
Let $1\leq \delta \leq k$, $k+\delta \leq n$ and $2\leq \alpha \leq q^k+1$ be integers.
\begin{enumerate}
    \item If $n<k+2\delta$, then
    $${\cal B}_q(n,k,\delta;\alpha) \geq (\alpha -1) q^{\max\{k,n-k\}(\min\{k,n-k\}-\delta+1)}.$$

    \item If $n\geq k+2\delta$, then for each $t$ such that $\delta\leq t\leq n-k-\delta$, we have \label{case:EK_2}
    \begin{enumerate}
        \item If $t<k$, then
        $${\cal B}_q(n,k,\delta;\alpha) \geq (\alpha-1)q^{k(t-\delta+1)}{\cal B}_q(n-t,k,\delta;\alpha).$$
        \item If $t\geq k$, then
        \begin{align*}{\cal B}_q(n,k,\delta;\alpha) \geq &(\alpha-1)q^{t(k-\delta+1)}{\cal B}_q(n-t,k,\delta;\alpha) \\
        &\ +{\cal B}_q(t+k-\delta,k,\delta;\alpha).
        \end{align*}
    \end{enumerate}
\end{enumerate}
\end{theorem}
Theorem~\ref{thm:EK_1_ext} improves the theorem above by removing the conditions $\alpha \leq q^k+1$ and $n<k+2\delta$.
For $n\geq k+2\delta$, the numerical results show that either could be better, depending on the parameters. The theoretical comparison between the two lower bounds is hard due to the recursive function and is left for {future research}.
\subsection{Discussion of Lower Bounds}
In the following, we compare the lower bound on $r_{\max}$ in Corollary~\ref{cor:EK19_lb} with the upper bounds in the previous sections.
\begin{itemize}
\item When $h \leq 2\ell$, Corollary~\ref{cor:EK19_lb} gives
$$r_{\max}\geq (\alpha-1)q^{\ell t(\eps t+1)},$$
which is close (up to a constant factor of $\gamma\approx 3.48$) to the upper bound in Corollary~\ref{cor:EZ19_vector}, i.e., $$r_{\max}< \gamma(\alpha-1)q^{\ell t (\eps t+1)}.$$
\item When $h \geq 2\ell$ and $\alpha =2$,
Corollary~\ref{cor:EK19_lb} gives
$$r_{\max}\geq q^{(h-\ell)(2h+\eps-h)t^2 +(h-\ell)t},$$
which is close (up to a constant factor of $\gamma$) to the upper bound in Theorem~\ref{thm:imupbound-2},  $$r_{\max}< \gamma q^{(h-\ell)(2h+\eps-h)t^2 +(h-\ell)t}.$$
\item The upper bound in Corollary~\ref{cor:imupperbound-N} cannot be applied here as $(h-\eps)/\ell \leq 2$.
\end{itemize}
{\subsection{Minimal Generalized Combination Networks}
Minimal networks are networks in which the removal of any single edge makes it unsolvable linearly (see~\cite{CCESW20} and the references therein). Thus, in such minimal networks, the number of messages equals the min-cut. These networks are of interest since they require the least amount of network resources to enable the multicast operation.

For generalized combination networks, minimality occurs exactly when $h=\alpha\ell+\varepsilon$. For the case $\alpha=2$, it has been shown in~\cite{EW18} that vector linear solutions do not outperform scalar linear solutions in minimal generalized combination networks. In Figures~\ref{fig:min-cut1} and~\ref{fig:min-cut2} we show two examples of our results for $\alpha>2$ in minimal generalized combination networks. The lower bound on the gap in Theorem~\ref{thm:gap_lb} may give a negative value, as illustrated in these two examples. This is because the gap between the necessary (blue curve) and the sufficient (green curve) conditions on $q^t$ such that a $(q,t)$-linear solution exists, is relatively large compared to non-minimal networks (cf.~Figures~\ref{fig:q_bound} and~\ref{fig:q_bound2}). Taking a closer look into the sufficient condition in Lemma~\ref{lem:ub_q}, the function $f(t)$ results in the different behaviors of the green curve for $h=\alpha\ell+\varepsilon$ and $h<\alpha\ell+\varepsilon$. We observe that $f(t)=\eps t+1$ if $h=\alpha\ell+\varepsilon$, and is quadratic in $t$ otherwise. Since the sufficient condition for $h\geq 2\ell+\eps$ is directly derived from the lower bound on $r_{\max}$ in Theorem~\ref{thm:LLL_bound}, one possible way to close the gap between the necessary and sufficient conditions is to improve the lower bound on the $r_{\max}$ in Theorem~\ref{thm:LLL_bound}, which we leave as an open question.
}
\begin{figure*}[t!]
  \centering
  \minipage{0.49\textwidth}
  \definecolor{darkgreen}{rgb}{0,0.7,0}
\begin{tikzpicture}
\pgfplotsset{compat = 1.3}
\begin{axis}[
	legend style={nodes={scale=0.7, transform shape}},
	width = \linewidth,
	height = \linewidth,
	title = {$h=8$, $\varepsilon=5$, $\ell=1$, $r=800000$, $\alpha=3$},
	xlabel = {{$t$}},
	xmin = 1,
	xmax = 19,
	ymin = 1,
	ymax = 2^12,
	legend pos = north east,
	legend cell align=left,
	ymode=log,
	grid=both,
yminorticks=true]

\addplot [solid, color=blue, thick, mark=*] table[row sep=\\] {
1.000000 7.826813 \\
2.000000 3.071920 \\
3.000000 2.163188 \\
4.000000 1.800155 \\
5.000000 1.607726 \\
6.000000 1.489196 \\
7.000000 1.409064 \\
8.000000 1.351352 \\
9.000000 1.307840 \\
10.000000 1.273877 \\
11.000000 1.246641 \\
12.000000 1.224316 \\
13.000000 1.205688 \\
14.000000 1.189910 \\
15.000000 1.176376 \\
16.000000 1.164639 \\
17.000000 1.154365 \\
18.000000 1.145296 \\
19.000000 1.137232 \\
};
\addlegendentry{{$\left(\frac{r+\theta-\alpha}{\gamma\cdot\theta}\right)^{\frac{1}{\ell (\varepsilon t+1)}}$ (Lemma~\ref{lem:lb_q})}};

\addplot [solid, color=darkgreen, thick, mark=diamond] table[row sep=\\] {
1.000000 162.129668 \\
2.000000 257.489238 \\
3.000000 306.263544 \\
4.000000 335.395425 \\
5.000000 354.683787 \\
6.000000 368.376508 \\
7.000000 378.593286 \\
8.000000 386.505865 \\
9.000000 392.813499 \\
10.000000 397.958904 \\
11.000000 402.235868 \\
12.000000 405.846950 \\
13.000000 408.936309 \\
14.000000 411.609322 \\
15.000000 413.944794 \\
16.000000 416.002832 \\
17.000000 417.830078 \\
18.000000 419.463279 \\
19.000000 420.931774 \\
};
\addlegendentry{{$\left(\frac{r}{\beta}\right)^{\frac{(\alpha-1)t}{f(t)}}$ (Lemma~\ref{lem:ub_q})}};

\addplot [solid, color=orange, thick, mark=star] table[row sep=\\] {
1.000000 2.000000 \\
2.000000 4.000000 \\
3.000000 8.000000 \\
4.000000 16.000000 \\
5.000000 32.000000 \\
6.000000 64.000000 \\
7.000000 128.000000 \\
8.000000 256.000000 \\
9.000000 512.000000 \\
10.000000 1024.000000 \\
11.000000 2048.000000 \\
12.000000 4096.000000 \\
13.000000 8192.000000 \\
14.000000 16384.000000 \\
15.000000 32768.000000 \\
16.000000 65536.000000 \\
17.000000 131072.000000 \\
18.000000 262144.000000 \\
19.000000 524288.000000 \\
};
\addlegendentry{{$2^t$}};

\addplot [dashed, color=black, thick, forget plot] table[row sep=\\] {
1.000000 162.129668 \\
5.000000 162.129668 \\
};
\addlegendentry{{$2^t$}};

\addplot [dashed, color=black, thick, forget plot] table[row sep=\\] {
1.000000 4.000000 \\
5.000000 4.000000 \\
};
\addlegendentry{{$2^t$}};

\draw[<->,>=latex, thick] (axis cs: 4.000000, 162.129668) -- node[right] {max $\gap(\cN) \approx 5.34$ bits} (axis cs: 4.000000, 4.000000);

\addplot [dashed, color=black, thick, forget plot] table[row sep=\\] {
1.000000 7.826813 \\
8.000000 7.826813 \\
};
\addlegendentry{{$2^t$}};

\addplot [dashed, color=black, thick, forget plot] table[row sep=\\] {
1.000000 512.000000 \\
8.000000 512.000000 \\
};
\addlegendentry{{$2^t$}};

\draw[<->,>=latex, thick] (axis cs: 7.000000, 7.826813) -- node[right] {min $\gap(\cN) \approx -6.03$ bits} (axis cs: 7.000000, 512.000000);

\end{axis}
\end{tikzpicture}
  \caption{{An illustration of the gap for a minimal generalized combination network $(5,1)-\mathcal{N}_{8,8e5,8}$, where the number of messages is the min-cut of the network.}}
  \label{fig:min-cut1}
  \endminipage\hfill
  \minipage{0.49\textwidth}
  \definecolor{darkgreen}{rgb}{0,0.7,0}
\begin{tikzpicture}
\pgfplotsset{compat = 1.3}
\begin{axis}[
	legend style={nodes={scale=0.7, transform shape}},
	width = \linewidth,
	height = \linewidth,
	title = {$h=13$, $\varepsilon=5$, $\ell=1$, $r=800000$, $\alpha=8$},
	xlabel = {{$t$}},
	xmin = 1,
	xmax = 79,
	ymin = 1,
	ymax = 2^34,
	legend pos = north east,
	legend cell align=left,
	ymode=log,
	grid=both,
yminorticks=true]

\addplot [solid, color=blue, thick, mark=*] table[row sep=\\] {
1.000000 7.826805 \\
2.000000 3.071919 \\
3.000000 2.163188 \\
4.000000 1.800155 \\
5.000000 1.607725 \\
6.000000 1.489196 \\
7.000000 1.409064 \\
8.000000 1.351352 \\
9.000000 1.307840 \\
10.000000 1.273877 \\
11.000000 1.246640 \\
12.000000 1.224316 \\
13.000000 1.205688 \\
14.000000 1.189910 \\
15.000000 1.176376 \\
16.000000 1.164639 \\
17.000000 1.154365 \\
18.000000 1.145296 \\
19.000000 1.137232 \\
20.000000 1.130015 \\
21.000000 1.123519 \\
22.000000 1.117640 \\
23.000000 1.112295 \\
24.000000 1.107414 \\
25.000000 1.102939 \\
26.000000 1.098822 \\
27.000000 1.095022 \\
28.000000 1.091503 \\
29.000000 1.088235 \\
30.000000 1.085192 \\
31.000000 1.082352 \\
32.000000 1.079695 \\
33.000000 1.077205 \\
34.000000 1.074865 \\
35.000000 1.072663 \\
36.000000 1.070586 \\
37.000000 1.068625 \\
38.000000 1.066770 \\
39.000000 1.065012 \\
40.000000 1.063345 \\
41.000000 1.061761 \\
42.000000 1.060254 \\
43.000000 1.058819 \\
44.000000 1.057451 \\
45.000000 1.056145 \\
46.000000 1.054897 \\
47.000000 1.053703 \\
48.000000 1.052560 \\
49.000000 1.051465 \\
50.000000 1.050414 \\
51.000000 1.049406 \\
52.000000 1.048437 \\
53.000000 1.047505 \\
54.000000 1.046608 \\
55.000000 1.045745 \\
56.000000 1.044913 \\
57.000000 1.044111 \\
58.000000 1.043337 \\
59.000000 1.042589 \\
60.000000 1.041867 \\
61.000000 1.041169 \\
62.000000 1.040494 \\
63.000000 1.039841 \\
64.000000 1.039208 \\
65.000000 1.038595 \\
66.000000 1.038001 \\
67.000000 1.037425 \\
68.000000 1.036867 \\
69.000000 1.036324 \\
70.000000 1.035798 \\
71.000000 1.035286 \\
72.000000 1.034789 \\
73.000000 1.034306 \\
74.000000 1.033836 \\
75.000000 1.033378 \\
76.000000 1.032933 \\
77.000000 1.032500 \\
78.000000 1.032077 \\
79.000000 1.031666 \\
};
\addlegendentry{{$\left(\frac{r+\theta-\alpha}{\gamma\cdot\theta}\right)^{\frac{1}{\ell (\varepsilon t+1)}}$ (Lemma~\ref{lem:lb_q})}};

\addplot [solid, color=darkgreen, thick, mark=diamond] table[row sep=\\] {
1.000000 4297331.400977 \\
2.000000 17227288.382877 \\
3.000000 28996708.500702 \\
4.000000 38088922.065522 \\
5.000000 45049603.934394 \\
6.000000 50474221.424259 \\
7.000000 54793871.634792 \\
8.000000 58303712.701942 \\
9.000000 61206655.102682 \\
10.000000 63644862.009668 \\
11.000000 65720164.362588 \\
12.000000 67507101.054501 \\
13.000000 69061340.011932 \\
14.000000 70425220.131596 \\
15.000000 71631475.583566 \\
16.000000 72705790.993298 \\
17.000000 73668589.251011 \\
18.000000 74536305.441760 \\
19.000000 75322310.005989 \\
20.000000 76037588.136926 \\
21.000000 76691246.931068 \\
22.000000 77290898.933207 \\
23.000000 77842955.706061 \\
24.000000 78352855.033775 \\
25.000000 78825238.571080 \\
26.000000 79264092.068978 \\
27.000000 79672857.038698 \\
28.000000 80054520.402341 \\
29.000000 80411687.021299 \\
30.000000 80746638.792496 \\
31.000000 81061383.122573 \\
32.000000 81357692.938987 \\
33.000000 81637139.910481 \\
34.000000 81901122.182616 \\
35.000000 82150887.655207 \\
36.000000 82387553.614844 \\
37.000000 82612123.370648 \\
38.000000 82825500.413133 \\
39.000000 83028500.515568 \\
40.000000 83221862.118064 \\
41.000000 83406255.271895 \\
42.000000 83582289.371466 \\
43.000000 83750519.861275 \\
44.000000 83911454.072841 \\
45.000000 84065556.320390 \\
46.000000 84213252.362765 \\
47.000000 84354933.321611 \\
48.000000 84490959.131521 \\
49.000000 84621661.586091 \\
50.000000 84747347.033983 \\
51.000000 84868298.771025 \\
52.000000 84984779.167564 \\
53.000000 85097031.564618 \\
54.000000 85205281.967613 \\
55.000000 85309740.562450 \\
56.000000 85410603.075286 \\
57.000000 85508051.994466 \\
58.000000 85602257.670666 \\
59.000000 85693379.309146 \\
60.000000 85781565.866270 \\
61.000000 85866956.860885 \\
62.000000 85949683.109852 \\
63.000000 86029867.395852 \\
64.000000 86107625.074660 \\
65.000000 86183064.628160 \\
66.000000 86256288.168691 \\
67.000000 86327391.899645 \\
68.000000 86396466.536675 \\
69.000000 86463597.693384 \\
70.000000 86528866.234958 \\
71.000000 86592348.602778 \\
72.000000 86654117.112786 \\
73.000000 86714240.230016 \\
74.000000 86772782.821503 \\
75.000000 86829806.389521 \\
76.000000 86885369.286911 \\
77.000000 86939526.916093 \\
78.000000 86992331.913169 \\
79.000000 87043834.318421 \\
};
\addlegendentry{{$\left(\frac{r}{\beta}\right)^{\frac{(\alpha-1)t}{f(t)}}$ (Lemma~\ref{lem:ub_q})}};

\addplot [solid, color=orange, thick, mark=star] table[row sep=\\] {
1.000000 2.000000 \\
2.000000 4.000000 \\
3.000000 8.000000 \\
4.000000 16.000000 \\
5.000000 32.000000 \\
6.000000 64.000000 \\
7.000000 128.000000 \\
8.000000 256.000000 \\
9.000000 512.000000 \\
10.000000 1024.000000 \\
11.000000 2048.000000 \\
12.000000 4096.000000 \\
13.000000 8192.000000 \\
14.000000 16384.000000 \\
15.000000 32768.000000 \\
16.000000 65536.000000 \\
17.000000 131072.000000 \\
18.000000 262144.000000 \\
19.000000 524288.000000 \\
20.000000 1048576.000000 \\
21.000000 2097152.000000 \\
22.000000 4194304.000000 \\
23.000000 8388608.000000 \\
24.000000 16777216.000000 \\
25.000000 33554432.000000 \\
26.000000 67108864.000000 \\
27.000000 134217728.000000 \\
28.000000 268435456.000000 \\
29.000000 536870912.000000 \\
30.000000 1073741824.000000 \\
31.000000 2147483648.000000 \\
32.000000 4294967296.000000 \\
33.000000 8589934592.000000 \\
34.000000 17179869184.000000 \\
35.000000 34359738368.000000 \\
36.000000 68719476736.000000 \\
37.000000 137438953472.000000 \\
38.000000 274877906944.000000 \\
39.000000 549755813888.000000 \\
40.000000 1099511627776.000000 \\
41.000000 2199023255552.000000 \\
42.000000 4398046511104.000000 \\
43.000000 8796093022208.000000 \\
44.000000 17592186044416.000000 \\
45.000000 35184372088832.000000 \\
46.000000 70368744177664.000000 \\
47.000000 140737488355328.000000 \\
48.000000 281474976710656.000000 \\
49.000000 562949953421312.000000 \\
50.000000 1125899906842624.000000 \\
51.000000 2251799813685248.000000 \\
52.000000 4503599627370496.000000 \\
53.000000 9007199254740992.000000 \\
54.000000 18014398509481984.000000 \\
55.000000 36028797018963968.000000 \\
56.000000 72057594037927936.000000 \\
57.000000 144115188075855872.000000 \\
58.000000 288230376151711744.000000 \\
59.000000 576460752303423488.000000 \\
60.000000 1152921504606846976.000000 \\
61.000000 2305843009213693952.000000 \\
62.000000 4611686018427387904.000000 \\
63.000000 9223372036854775808.000000 \\
64.000000 18446744073709551616.000000 \\
65.000000 36893488147419103232.000000 \\
66.000000 73786976294838206464.000000 \\
67.000000 147573952589676412928.000000 \\
68.000000 295147905179352825856.000000 \\
69.000000 590295810358705651712.000000 \\
70.000000 1180591620717411303424.000000 \\
71.000000 2361183241434822606848.000000 \\
72.000000 4722366482869645213696.000000 \\
73.000000 9444732965739290427392.000000 \\
74.000000 18889465931478580854784.000000 \\
75.000000 37778931862957161709568.000000 \\
76.000000 75557863725914323419136.000000 \\
77.000000 151115727451828646838272.000000 \\
78.000000 302231454903657293676544.000000 \\
79.000000 604462909807314587353088.000000 \\
};
\addlegendentry{{$2^t$}};

\addplot [dashed, color=black, thick, forget plot] table[row sep=\\] {
1.000000 4297331.400977 \\
5.000000 4297331.400977 \\
};
\addlegendentry{{$2^t$}};

\addplot [dashed, color=black, thick, forget plot] table[row sep=\\] {
1.000000 4.000000 \\
5.000000 4.000000 \\
};
\addlegendentry{{$2^t$}};

\draw[<->,>=latex, thick] (axis cs: 4.000000, 4297331.400977) -- node[right] {max $\gap(\cN) \approx 20.04$ bits} (axis cs: 4.000000, 4.000000);

\addplot [dashed, color=black, thick, forget plot] table[row sep=\\] {
1.000000 7.826805 \\
31.000000 7.826805 \\
};
\addlegendentry{{$2^t$}};

\addplot [dashed, color=black, thick, forget plot] table[row sep=\\] {
1.000000 134217728.000000 \\
31.000000 134217728.000000 \\
};
\addlegendentry{{$2^t$}};

\draw[<->,>=latex, thick] (axis cs: 30.000000, 7.826805) -- node[right] {min $\gap(\cN) \approx -24.03$ bits} (axis cs: 30.000000, 134217728.000000);

\end{axis}
\end{tikzpicture}
  \caption{{An illustration of the gap for a minimal generalized combination network $(5,1)-\mathcal{N}_{13,8e5,13}$, where the number of messages is the min-cut of the network.}}
  \label{fig:min-cut2}
  \endminipage
\end{figure*}

\section{Conclusion}
In this work, we studied necessary and sufficient conditions for the existence of $(q,t)$-linear solutions to the generalized combination network $(\eps,\ell)$-$\mathcal{N}_{h,r,\alpha\ell+\eps}$. We derived new upper and lower bounds on $r_{\max}$, the maximum number of nodes in the middle layer, for a fixed network structure, thus getting bounds on the field size of the scalar/vector solution. Our lower bound is close (within a constant factor of $\gamma\approx 3.48$) to our upper bound for $h\leq 2\ell$ or $h\geq 2\ell,\alpha=2$. We summarize the best known bounds on $r_{\max}$ for different parameter ranges, in Table~\ref{tab:r_bound_summary}.

\begin{table*}[t!]
  \caption{Bounds on $r_{\max}$ of the $(\eps,\ell)-\cN_{\alpha,r,\alpha\ell+\eps}$ network with $(q,t)$-linear solutions}
  \label{tab:r_bound_summary}
  \begin{center}
  \begin{tabular}[l]{|l|l|l|l|l|}
    \hline
    Upper Bounds & $h< 2\ell+\eps$& Reference &$h\geq 2\ell+\eps$ & Reference\\
    \hline
    $\alpha> 2$ & $r_{\max}<\gamma (\alpha-1)q^{\ell t(\eps t+1)}$ &\cite{EZjul2019} (see also Corollary~\ref{cor:EZ19_vector}) &$r_{\max}<\gamma\theta q^{\ell t(\eps t+1)}+\alpha-\theta$  & Corollary~\ref{cor:imupperbound-N} \\
    \hline
    \multirow{2}{*}{$\alpha=2$}&\multicolumn{3}{c}{\multirow{2}{*}{$r_{\max}<\gamma q^{\min \{\ell t(\eps t+1),(h-\ell)(2\ell+\eps-h)t^2+(h-\ell)t\}}$}}& \cite{EZjul2019} \& Corollary~\ref{cor:imupperbound-2-Network}\\ &\multicolumn{3}{c}{\ }& (Comparison in Lemma~\ref{lem:compare-alpha2}) \\
    \hline
    \hline
    Lower Bounds& $h< 2\ell+\eps$& Reference & $h\geq 2\ell+\eps$ & Reference\\
    \hline
    $\alpha\geq 2$ & $r_{\max}\geq (\alpha-1)q^{g(t)}$& Corollary~\ref{cor:EK19_lb}& $r_{\max}\geq \beta \cdot q^{\frac{f(t)}{\alpha-1}}$ &  Theorem~\ref{thm:LLL_bound} \\
    \hline
  \end{tabular}
  \end{center}
  Remarks: The bounds are valid for $\alpha\geq 2,h,\ell\geq 1,\eps\geq 0$. For non-trivially solvable generalized combination networks, $\ell+\eps\leq h\leq \alpha\ell+\eps$. The other parameters are $\gamma\approx3.48$, $\beta = \parenv*{(\alpha-1)!/(2e\gamma\alpha)}^{1/(\alpha-1)}$,  $f(t)=(\alpha\ell+\eps-h)\eps t^2+(\alpha\ell+2\eps-h)t +{1}$, $\theta = \alpha-\floor*{(h-\eps)/\ell}+1$, and  $g(t)=\max\{\ell t,(h-\ell)t\}\cdot(\min\{\ell t, (h-\ell)t\}-(h-\ell-\eps)t+1)$.
\end{table*}

Moreover, we studied the gap between the minimal field size of a scalar solution and a vector solution.
{For general multicast networks, the gap is not well defined, since some multicast networks with no scalar $(q_s,1)$-linear solution, but only vector $(q,t)$-linear solutions (even for $q_s>q^t$), were demonstrated in~\cite{SYL+dec2016}. In our work, we focused on a class of multicast networks, the generalized combination networks, where both scalar and vector solutions exist. We studied the gap for this specific class of multicast networks. Unlike the previous works~\cite{EW18,CCESW20},} which focused on engineering the networks to obtain a high gap, we started by fixing network parameters (i.e., $h,r,\alpha,\ell,\eps $), and then provided bounds for its gap, which do not depend on $t$. Of particular interest is the conclusion from~Corollary~\ref{cor:gap_ub} and Corollary~\ref{cor:gap}: fixing the number of messages $h$, and parameters relating to the connectivity level of the network (i.e., $\alpha,\ell,\eps$), we only vary the number of middle layer nodes, $r$, or equivalently, the number of receivers $N:=\binom{r}{\alpha}$, proving that the gap is $\gap(\cN)=\Theta(\log r)=\Theta(\log N)$. Namely, the scalar linear solutions over-pays an order of $\log(r)$ extra bits per symbol to solve the network, in comparison to the vector linear solutions.

The novel upper and lower bounds on the gap cover all generalized network parameters, except $\eps=0$. This may imply that the direct links between the source and the terminals are crucial for vector network coding to have an advantage in generalized combination networks. The direct link in usual communication networks might not be practical, however, in some recent applications, such as coded caching, this direct link can be seen as the cached content at the receivers. The exact nature of the connection between direct links and field-size gap, is left for future work.

\bibliographystyle{IEEEtranS}
\bibliography{VNC}

\begin{IEEEbiographynophoto}{Hedongliang Liu}
  (S'19) is a doctoral student at the Technical University of Munich (TUM), Germany in the Department of Electrical and Computer Engineering.
  She received the B.Eng.~degree in Information Engineering at Southeast University, Nanjing, P.~R.~China in 2016 and M.Sc.~degree in Communication Engineering in 2019.
  Her research interests are coding theory and its application in networks, storage, and cryptography.
\end{IEEEbiographynophoto}

\begin{IEEEbiographynophoto}{Hengjia Wei}
  is a postdoctoral fellow in the School of Electrical and Computer Engineering, Ben-Gurion University of the Negev, Israel.
  He received the Ph.D.~degree in Applied Mathematics from Zhejiang University, Hangzhou, Zhejiang, P.~R.~China in 2014. He was a postdoctoral fellow in the Capital Normal University, Beijing, P.~R.~China, from 2014 to 2016, and a research fellow in the School of Physical and Mathematical Sciences, Nanyang Technological University, Singapore, from 2016 to 2019.
  His research interests include combinatorial design theory, coding theory and their intersections.

  Dr. Wei received the 2017 Kirkman Medal from the Institute of Combinatorics and its Applications.
\end{IEEEbiographynophoto}
\begin{IEEEbiographynophoto}{Sven Puchinger}
  (S'14, M'19) is a postdoctoral researcher at the Technical University of Munich (TUM), Germany.
  He received the B.Sc.~degree in electrical engineering and the B.Sc.~degree in mathematics from Ulm University, Germany, in 2012 and 2016, respectively. During his studies, he spent two semesters at the University of Toronto, Canada. He received his Ph.D.~degree from the Institute of Communications Engineering, Ulm University, Germany, in 2018. He has been a postdoc at the Technical University of Munich (2018--2019 and since 2021) and the Technical University of Denmark (2019--2021), Denmark.
  His research interests are coding theory, its applications, and related computer-algebra methods.
\end{IEEEbiographynophoto}

\begin{IEEEbiographynophoto}{Antonia Wachter-Zeh}
  (S’10–M’14-SM'20) is an Associate Professor at the Technical University of Munich (TUM), Munich, Germany in the Department of Electrical
  and Computer Engineering.
  She received the M.Sc.~degree in communications technology in 2009 from Ulm University, Germany. She obtained her Ph.D.~degree in 2013 from Ulm University and from Universite de Rennes 1, Rennes, France. From 2013 to 2016, she was a postdoctoral researcher at the Technion—Israel Institute of Technology, Haifa, Israel, and from 2016 to 2020 a Tenure Track Assistant Professor at TUM.
  Her research interests are coding theory, cryptography and information theory and their application to storage, communications, privacy, and security.

  Prof.~Wachter-Zeh is a recipient of the DFG Heinz Maier-Leibnitz-Preis and of an ERC Starting Grant.
\end{IEEEbiographynophoto}

\begin{IEEEbiographynophoto}{Moshe Schwartz}
(M'03--SM'10)
is a professor at the School of Electrical and Computer
Engineering, Ben-Gurion University of the Negev, Israel. His research
interests include algebraic coding, combinatorial structures, and
digital sequences.

He received the B.A.~(\emph{summa cum laude}), M.Sc., and
Ph.D.~degrees from the Technion -- Israel Institute of Technology,
Haifa, Israel, in 1997, 1998, and 2004 respectively, all from the
Computer Science Department. He was a Fulbright post-doctoral
researcher in the Department of Electrical and Computer Engineering,
University of California San Diego, and a post-doctoral researcher in
the Department of Electrical Engineering, California Institute of
Technology. While on sabbatical 2012--2014, he was a visiting scientist
at the Massachusetts Institute of Technology (MIT).

Prof.~Schwartz received the 2009 IEEE Communications Society Best
Paper Award in Signal Processing and Coding for Data Storage, and the
2020 NVMW Persistent Impact Prize. He has also been serving as an
Associate Editor for Coding Techniques for the IEEE Transactions on
Information Theory since 2014, and an Editorial Board Member for the
Journal of Combinatorial Theory Series A since 2021.
\end{IEEEbiographynophoto}
\end{document}